\documentclass[journal,12pt,draftclsnofoot,onecolumn]{IEEEtran}
\usepackage{bm} 
\usepackage{multicol}     
\usepackage{multirow}
\usepackage{booktabs}
\usepackage{bigstrut}
\usepackage{hyperref}
\usepackage[table]{xcolor}
\usepackage{color}
\usepackage{balance}
\usepackage{graphicx}
\usepackage{enumerate}
\usepackage{subfigure}
\usepackage{algorithm}
\usepackage{algorithmic}
\usepackage{booktabs}
\usepackage{bbold}
\usepackage{cite}
\usepackage{tcolorbox}
\usepackage{setspace}
\usepackage{amsmath}
\usepackage{amssymb}
\usepackage{textcomp}
\usepackage{amsthm}
\usepackage{authblk}
\usepackage[utf8]{inputenc}
\usepackage{amsfonts,amssymb}
\theoremstyle{remark}

\begin{document}
 \title{Active RIS-aided EH-NOMA Networks: A Deep Reinforcement Learning  Approach }
\author{ Zhaoyuan Shi,    Huabing Lu, Xianzhong Xie, Helin Yang, Chongwen Huang, {\textit{Member, IEEE}}, Jun Cai, {\textit{Senior Member, IEEE}}, and Zhiguo Ding,  {\textit{Fellow, IEEE}}
	
\thanks{ 
%
Z. Shi is with the Key Laboratory of Intelligent Perception and Computing of Anhui Province, Anqing Normal University, Anqing,  China (e-mail:shizy123@126.com).

H. Lu is with the Key Laboratory of Intelligent Control and Optimization for Industrial Equipment of Ministry of Education, Dalian University of Technology, Dalian 116024, China (e-mail: luhuabing@dlut.edu.cn)

X. Xie is  with Chongqing Key Lab of Computer networks and Communication Technology, Chongqing University of Posts and Telecommunications, Chongqing, China(email:xiexzh@cqupt.edu.cn).
	
H. Yang is with the Department of Information and Communication Engineering, School of Informatics, Xiamen University, Xiamen 361005, China (e-mail: helinyang066@xmu.edu.cn).
   
C. Huang is with the College of Information Science and Electronic Engineering, Zhejiang University, Hangzhou 310027, China. (e-mail: chongwenhuang@zju.edu.cn)
	
	
	J. Cai is with the Network Intelligence and Innovation Lab (NI$^2$L), Department of Electrical and Computer Engineering, Concordia University, Montreal, QC H3G 1M8, Canada (e-mail: jun.cai@concordia.ca).
	
	Z. Ding is with the Department of Electrical Engineering, Princeton University, Princeton, NJ 08544, USA. He is also with the School of
	Electrical and Electronic Engineering, the University of Manchester, Manchester, UK (e-mail: zhiguo.ding@manchester.ac.uk).
}
}
\maketitle
\vspace{-30pt}
\begin{abstract}

An active reconfigurable  intelligent surface (RIS)-aided
multi-user downlink communication system  is investigated, where   non-orthogonal multiple
access (NOMA)  is employed to improve spectral efficiency, and the active RIS    is powered by energy harvesting (EH). The problem of joint  control of the  RIS's amplification matrix and phase shift matrix is formulated to maximize the communication success ratio with  considering the
quality of service (QoS) requirements of users, dynamic communication state, and dynamic available energy of RIS. To tackle this non-convex problem, a cascaded deep learning algorithm namely long
short-term memory-deep deterministic policy gradient  (LSTM-DDPG)  is designed. First, an advanced LSTM based algorithm is developed to predict users’ dynamic communication state. Then, based on the  prediction results,  a
DDPG based algorithm is proposed to joint control the amplification matrix and phase shift matrix of the RIS. Finally,    simulation results verify the accuracy of the prediction of the proposed LSTM algorithm, and demonstrate that the LSTM-DDPG algorithm has a significant advantage over other benchmark algorithms in terms of communication success ratio performance.

\end{abstract}

\IEEEpeerreviewmaketitle
\section{Introduction}

\renewcommand{\baselinestretch}{0.75} 

\IEEEPARstart{W}ith the ability to actively reconfigure the wireless communication environments, the reconfigurable intelligent surfaces (RIS), also named  intelligent reflecting surfaces (IRS), has become a focal point in the field of wireless communications \cite{RIS,RIS_suv}. By controlling low-cost passive components, RIS  can provide  users with an additional set of cascaded channels in addition to the direct link, thus effectively improving communication performance.  The RIS   is a planar array of a large number of passive elements that are reconfigurable and capable of reflecting electromagnetic signals in the desired manner (controlled by an attached intelligent	RIS controller)\cite{RIS1}. Compared with other candidate technologies (such as active relays), RIS   has significant advantages in terms of energy consumption, flexible deployment, negligible noise, and economic cost. Therefore, RIS   has been widely recognized as a promising paradigm for future 6-th Generation Mobile Communication (6G) networks \cite{RIS2}.

The significant capacity gain that RIS   brings to wireless communications is mainly derived from negligible noise on the RIS  cascade channel \cite{RIS3}. The array gain obtained by an RIS with  $M$ elements is proportional to $M^2$, which is $M$ times more than that achievable by a multiple-input multiple-output (MIMO) system with $M$ antennas at the base station (BS)
 \cite{active_ris1}. However, to obtain such capacity gains, it is usually assumed that the quality of the direct channel (from the transmitter to the receiver) in the system is very poor, or even completely blocked \cite{RIS2,RIS3,RIS4,RIS5,RIS6}. Otherwise, the gain of RIS is insignificant or even negligible. The reason behind this phenomenon is the multiplicative fading effect in the cascade channel. Specifically, the path loss of the cascaded channel is the multiplication of the path loss on the BS-RIS link and the RIS-user link, which is usually thousands of times larger than that  on the direct channel \cite{RIS7}. Clearly, the multiplicative fading effect reduces the capacity gain brought by RIS    arrays and limits the application scenarios of RIS. Therefore, most of the existing RIS-related studies have bypassed this effect and only considered the case where the quality  of direct channel is poor or blocked \cite{RIS2,RIS3,RIS4,RIS5,RIS6}.

In order to overcome fundamental performance bottleneck brought by the multiplicative fading effect in the RIS, a novel concept called active RIS    was proposed \cite{active_ris1}. The main concept of active RIS  is to integrate a power amplifier in each RIS element, thus the RIS can actively amplify the reflected signal. The introduction of active power amplifiers allows RIS to achieve a sizable increase in capacity, regardless of whether the direct channel is poor or not \cite{active_ris}.    

 Because of the performance advantages, active RIS has attracted a lot of  attention  in recent years \cite{active_ris1,active_ris,active_ris3,active_ris4,active_ris5}. Zhang \textit{et al.} developed an active RIS    model in \cite{active_ris1}, which was validated through experimental measurements on a fabricated active RIS element. Based on the model, the authors analyzed the asymptotic performance of active RIS. Finally, a joint transmit beamforming and reflect precoding algorithm was proposed for maximizing sum rate of the active RIS-aided MIMO system.  
To reduce the energy consumption of active RIS, Liu \textit{et al.} proposed a sub-connected architecture   \cite{active_ris}, i.e., some RIS    elements control their phase shifts independently but
sharing the same power amplifier. Furthermore, a beamforming algorithm with the architecture was developed to maximize the energy efficiency of the active RIS-aided system. In \cite{active_ris3}, the authors  fundamentally proved that the multiplicative fading can be transformed into additive fading in active RIS. 
In \cite{active_ris4}, You  \textit{et al.}  defined that in passive RIS, all power amplification factors are the same and $\eta_m=1,\forall m \in \mathcal{M}$;  while in active RIS, each power amplification factor is larger than one due to the introduction of an active amplifier. Besides, the simulation results indicate that active RIS    can perform better than passive RIS    under optimized placement in most practical scenarios.  The work in \cite{active_ris5} theoretically compared the active RIS     with the passive RIS     under the same overall power budget. The theoretical and numerical results showed that the active RIS    is superior to passive RIS    when the power budget is not very small.

It has been proven that active RIS    usually has significant performance advantages over passive RIS. These, however, come at higher energy consumption, which is contrary to the green philosophy of low cost and low energy consumption of RIS. Although \cite{active_ris} proposed a novel energy-efficient structure, it still requires a fixed power supply to provide more energy than the passive RIS. Thanks to the rapid development of green energy technology in recent years \cite{eh_survey} and the geographical advantages of RIS    installation locations, energy harvesting (EH) technology  has naturally become our first choice to address the energy supply problem in active RIS.  

\textcolor{black}{In addition, as a promising candidate technology for 6G  networks , NOMA technology is effective in improving spectrum efficiency and increasing user connectivity \cite{NOMA1}. Therefore,  the NOMA technology is introduced  in this paper.}
 NOMA allows multiple users of different power levels to communicate simultaneously with the same frequency/time/code resources, and separates the multi-user signals by applying successive interference cancellation (SIC) at the receivers \cite{RIS_noma}.  \textcolor{black}{In  \cite{Wang}, Wang \textit{et al.}   combines NOMA with EH techniques and applies them to UAV communication systems. The analytical and simulation results show that the proposed scheme can substantially outperform the alternatives that do not use NOMA and EH techniques.}

\textcolor{black}{The combination of  NOMA, EH, and RIS     can effectively address some challenges of 6G, including spectrum efficiency improvement, energy consumption reduction, and system capacity improvement.}  Currently, Alanazi  has studied RIS-aided EH-NOMA  system under Nakagami  \cite{ris_eh_noma1} and Rayleigh fading channels \cite{ris_eh_noma2}. In both channels, the transmitter harvested radio frequency (RF) energy from another node, which is used to transmit data to multiple NOMA users by using RIS. \textcolor{black}{ However, in both papers, each user receives the same phase signal through the RIS. Moreover, the phase and power of RIS were not optimized in these two papers. }
 Diamanti \textit{et al.} studied the problem of maximizing the uplink and downlink rates of Internet of Things (IoT) users in an RIS-aided EH-NOMA system \cite{ris_eh_noma3}, where the IoT users are simultaneous wireless information and power transfer (SWIPT) nodes that receive signals while harvesting energy for their uplink and downlink communications. \textcolor{black}{Zhang \textit{et al.} \cite{Zhang} proposed a RIS-aided cooperative transmission scheme using hybrid SWIPT and transmit antenna selection (TAS) protocols. The simulation work demonstrates the performance advantages brought by the combination of EH, RIS, and NOMA.} 
To optimize the objective, an iterative algorithm was proposed for jointly optimizing the phase of the RIS elements and the power allocation of the IoT users.  Unlike the existing literature where the users are the EH node, in this paper the EH technique is employed at the RIS, and the harvested energy is used for the amplification and transmitting of the arriving signals.


In this paper, we  focus on the joint control of  amplification matrix and the phase shift matrix of the active RIS to maximize the communication success ratio of the RIS-aided EH-NOMA networks, while satisfying the  quality of service (QoS) requirement of users, dynamic communication state,  and the dynamic energy constraint at the RIS. 
However, due to the stochastic nature of harvested energy, communication state, and the wireless channel, the traditional optimization algorithms are hardly applicable anymore. Thanks to the rapid development of artificial intelligence, the paradigm of deep reinforcement learning (DRL) offers a promising approach \cite{duibi_ding,ziji}. Due to the powerful learning capability in dynamic unknown environments, DRL has been widely applied to learn the optimal decision policy in wireless communications  \cite{DRL_SUV}.

In recent years, the DRL algorithms have been studied in the  RIS    \cite {ris_drl1,ris_drl2,ris_drl3,ris_drl4,ris_drl5,SWAGN}.
Faisal \textit{et al.} investigated an RIS    aided full-duplex multiple-input-single-output (MISO) wireless system, where the beamforming  and the RIS    phase shift were designed for maximizing the sum rate. A novel DRL algorithm was proposed for optimizing the RIS phase shift \cite{ris_drl1}.  In \cite{ris_drl2}, Liu \textit{et al.} designed a novel  double deep
Q-network (D3QN) based position-acquisition and phase-control algorithm for the RIS-aided MISO-NOMA network, where an BS periodically observes the  environment state   for  optimizing the deployment and phase shift of the RIS    by learning from its mistakes and the feedback of users. In \cite{ris_drl3},  Mu \textit{et al.} investigated the optimization problem of maximizing the effective throughput by optimizing the phase shift of the RIS  and the power allocation of the BS in RIS-aided networks by deep learning (DL) and reinforcement learning (RL) algorithms, respectively, where the NOMA and orthogonal multiple access (OMA) schemes were both employed. The simulation results showed that in the RIS-aided network, NOMA achieves a 42\% gain compared to OMA, and the RL algorithm also achieves better communication performance than the DL algorithm. In \cite{ris_drl4}, an RIS-aided unmanned aerial vehicle (UAV)-NOMA system was studied, and a DRL algorithm was proposed for the UAV trajectory, RIS configuration, and power control. In \cite{ris_drl5}, a DRL based phase shift optimization algorithm was proposed for RIS    aided MISO system, where both half-duplex (HD) and full-duplex (FD) modes were considered. \textcolor{black}{ In \cite{SWAGN}, Wang \textit{et al.} presented a novel and effective DRL-based approach to address joint resource management  in a practical multi-carrier NOMA system.}

The above DRL based optimization algorithms have achieved good performance in RIS-aided networks. However, these algorithms are based on some perfect assumptions, such as  that the agent is accurately informed of the perfect channel state information for all channels \cite{ris_drl2,ris_drl3,ris_drl4} and  that  users are always in the communication state \cite{ris_drl1,ris_drl3,ris_drl4,ris_drl5,SWAGN}. In addition, the states in these DRL algorithms usually contain the phase shift of all RIS    elements and (or) all channel gains \cite{ris_drl1, ris_drl2,ris_drl3,ris_drl5}. To obtain this information, it is often assumed that a wired link is erected between the RIS controller and the BS, which undoubtedly imposes an additional cost overhead on the RIS-assisted wireless network.  Further more, such a DRL state setting always leads to a high complexity of the algorithm, as the number of RIS components in the system is usually large. In addition, to the best of our knowledge, there is no research related to the DRL algorithm for active RIS-aided EH-NOMA networks.


Motivated by the aforementioned background, we design a long short-term memory-deep deterministic policy gradient  (LSTM-DDPG) algorithm to control the RIS, which is composed of the LSTM network and the DDPG network. The LSTM network is used to predict the environment features i.e., the user's communication state (UCS), and the DDPG framework is designed to make controlling decisions based on the prediction result of LSTM. The main contributions of our work are summarized as follows:

\begin{itemize}
\item 	
An innovative framework for active RIS-aided EH-NOMA networks is proposed,  in which the UCS is dynamically changing and the active RIS is powered by the EH. Besides, the wired connection between the RIS controller and the BS is no longer needed. Based on the framework, we formulate an optimization problem with the objective of maximizing the communication success ratio by jointly controlling the amplification matrix and phase shift matrix of the active RIS.

\item  A low-complexity LSTM based algorithm is developed for UCS prediction. The prediction is formulated as a time series prediction problem, and the algorithm is trained based on an empirical dataset of long-term observations. After the training, the LSTM network can predict the current communication state information based on the historical UCS.

\item  A  novel DDPG based algorithm is proposed for the joint control of the amplification power matrix and the phase shift matrix of the active RIS. Unlike existing DRL algorithms that require high-dimensional RIS    channel and phase shift information, this algorithm takes the prediction results of the LSTM and the currently available energy of the RIS as input states.  Besides,  as an agent, the RIS controller   does not require additional information interaction with the BS or the users throughout the learning process of the algorithm.

\item The complexity of the LSTM based UCS prediction algorithm and the DDPG based RIS  control algorithm are analyzed. Extensive simulation are provided to verify the high prediction accuracy of the LSTM algorithm, and also demonstrate that the proposed
DDPG based RIS control algorithm outperforms the existing benchmarks in terms of the communication
success ratio.

	
\end{itemize}

The remainder of this paper is structured as follows. The system model is described in Section \uppercase\expandafter{\romannumeral2}. In Section \uppercase\expandafter{\romannumeral3}, LSTM based UCS prediction algorithm is presented. Section \uppercase\expandafter{\romannumeral4} presents the details of the proposed LSTM-DDPG algorithm. Simulation results and concluding remarks are finally presented in Sections \uppercase\expandafter{\romannumeral5} and \uppercase\expandafter{\romannumeral6}, respectively.

\section{System Model}
\begin{figure}[t]
	\centering
	\includegraphics[width=6.5in]{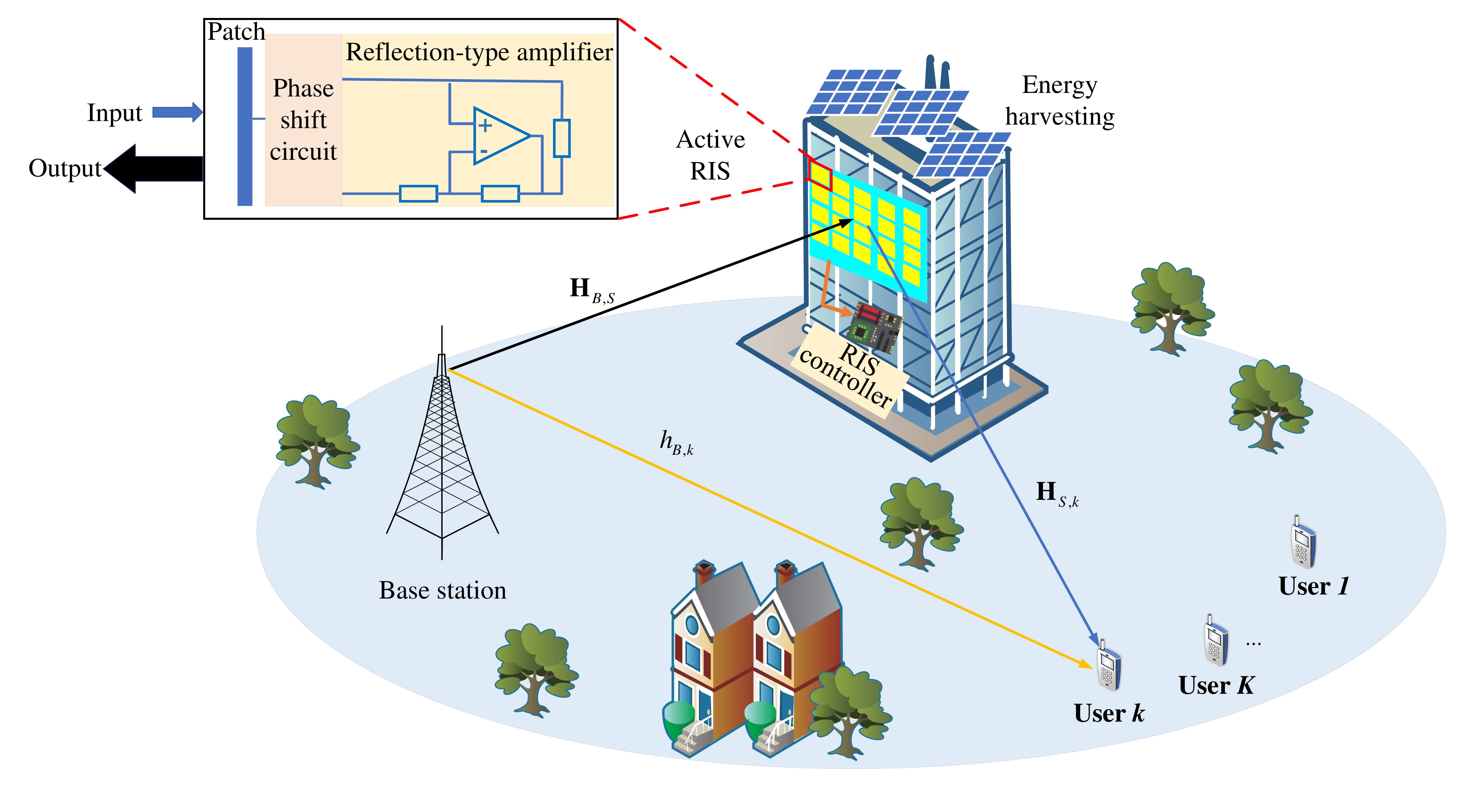}
		\caption{Illustration of  active RIS-aided    EH-NOMA networks. }
	\centering
	\label{Fig1}	
\end{figure}

As illustrated in Fig. \ref{Fig1}, an active RIS-aided EH-NOMA communication between a  BS and $K$  users is considered. As in \cite{ris_drl3,ris_drl4}, the BS and users are all equipped with single antenna, and the locations of the BS, RIS are fixed, which are  denoted in a three-dimensional (3D) Cartesian coordinate system.  The communication is aided by an active RIS with $M$ active reflecting elements that are installed on the facade of a building \cite{RIS_suv}. EH  is introduced to provide green energy for the active RIS, and NOMA is invoked to further improve the spectrum efficiency of the network \cite{RIS_noma}. The  amplification matrix and phase shift matrix of active RIS are controlled by an attached smart RIS controller.

\subsection{ Active RIS  Model}
 
Similar to existing passive RIS, active RIS   can reflect the incident signal by reconfiguring the phase shift. The most essential difference between them is that passive RIS only reflects but does not amplify the incident signal, while active RIS    can further amplify the reflected signal \cite{active_ris,active_ris1}. To amplify the signal, an additional amplification device needs to be integrated on the passive RIS as shown in Fig. \ref{Fig1}.


Note that there is a clear difference between active RIS and relay-type RIS. In the relay-type RIS, numerous passive RIS elements are connected to an active RF chain and therefore it has some signal processing capability and can transmit the pilot signal independently.
However, active RIS    is simple in structure and only uses  power amplifiers and phase shifting circuits to control the signal, without  signal processing and analysis capability, thus offering significant advantages in terms of delay and cost effectiveness \cite{active_ris1,active_ris}.

The key component of an active RIS   element is the
additionally integrated active reflection-type amplifier, which
can be implemented  by the existing active components, such as current-inverting converters or some integrated circuits \cite{active_ris}. With reflection-type amplifiers supported by a power supply, the reflected and amplified signal of an $M$-elements active RIS    can be modeled as follows:
\begin{equation}
\setlength{\abovedisplayskip}{3pt}
\setlength{\belowdisplayskip}{3pt}
\mathbf{y}_S(t)=\underbrace{\mathbf{P(t)}\mathbf{\Theta(t) x(t)}}_{\text{desired signal}}+\underbrace{\mathbf{P(t)\Theta(t) z}}_{\text{dynamic noise}}+\underbrace{\mathbf{n}_s}_{\text{static noise}},
\label{m1}
\end{equation}
where $\mathbf{P(t)}=\text{diag}\left(p_1^S(t),p_2^S(t),\cdots,p_M^S(t)\right)\in\mathbb{R}_+^{M\times M}$ is the amplification matrix, wherein each element $p_m^S(t)$ is the amplification factor for the $m$-th RIS    element at slot $t$ (where the superscript $S$ refers to the RIS), and $\textbf{x}\in \mathbb{C}^{M}$ is the incident signal. Each element of $\textbf{P}(t)$ satisfies $0\leq p_m^S\leq L$, where $L$ is the maximum  amplification  factor.
The $\bm{\Theta(t)}$ denotes the phase shift matrix, which can be expressed by 
\begin{equation}
\mathbf{\Theta(t)}=\text{diag}(e^{j\theta_1(t)},e^{j\theta_2(t)},\cdots,e^{j\theta_{M}(t)}),
\end{equation}
where $\theta_m(t)\in[0  ,2\pi)$ is the reflection phase shift of the $m$-th RIS    element.
Due to the fact that active components are required for signal amplification in active RIS, the thermal noise introduced by active components cannot be ignored. As shown in (\ref{m1}), the noise  introduced by the active RIS   contains dynamic noise $\mathbf{P(t)\Theta(t) z}$ and static noise $\textbf{n}_s$, but only $\mathbf{P(t)\Theta(t) z}$ is associated with the amplification matrix $\mathbf{P}$, and $\mathbf{z}\sim \mathcal{CN}(\mathbf{0}_M,\delta_z^2\mathbf{I}_M)$. Therefore, compared with dynamic noise, $\mathbf{n}_s$ can be almost ignored.

\textbf{Remark 1.} \textit{Unlike the definition of the active RIS in  \cite{active_ris4} with respect to the amplification factor(i.e., $p_m^S>1$), the value range of $p_m^S$ in this paper is $[0,L]$. This is because the active RIS in this paper is powered by energy harvesting and therefore the harvested energy is dynamic, which makes it difficult to guarantee that the amplification factor of all RIS elements is greater than 1.}

\textcolor{black}{The positions of the BS, RIS, and users are modeled in a three-dimensional (3D) Cartesian coordinate system. Let $\textbf{H}_{B,S}(t)\in\mathbb{C}^{M\times 1}$ and  ${h}_{B,k}(t)\in\mathbb{C}^{1\times 1}$ denote the BS-RIS    and BS-user $k$ channels at $t$-th time slot, respectively. Correspondingly, $\textbf{H}_{S,k}\in\mathbb{C}^{1\times M}$ denotes the channel between the RIS    to user $k$. It is assumed that all channels obey a static block fading model in which the channel parameters remain  constant within each time slot and vary independently in the next time slot. Considering that the RIS-assisted communication environment is always scattered and highly correlated, this paper adopts the channel model  in \cite{channel}. Specifically, the direct link ${h}_{B,k}(t)$ follows Rayleigh fading distribution. The channels $\textbf{H}_{S,k}$ and $\textbf{H}_{B,S}$ are defined as 
\begin{equation}
\begin{aligned}
&\textbf{H}_{S,k}\in\mathcal{N}_{\mathbb{C}}(\mathbf{0},A\mu_{S,k}\mathbf{R}),\\
&\textbf{H}_{B,S}\in\mathcal{N}_{\mathbb{C}}(\mathbf{0},A\mu_{B,S}\mathbf{R}),
\end{aligned}
\end{equation}
where $A$ is the area of each RIS element. $\mu_{S,k}$ and $\mu_{B,S}$ are the average large scale path loss  of channel $\textbf{H}_{S,k}$ and channel $\textbf{H}_{B,S}$, respectively. The  path loss is calculated as $C_0(\dfrac{d}{d_0})^{-\alpha}$ \cite{RIS}, where $C_0$ denotes the pass loss  in the condition of  the reference distance $d_0=1$ meter (m), $d (d_{B,k},d_{B,S},d_{S,k})$ represents  the link distance, $\alpha (\alpha_{B,k},\alpha_{B,S},\alpha_{S,k}) $ denotes the path loss exponent. $\mathbf{R}\in\mathbb{C}^{M\times M}$ represents the normalized spatial correlation matrix which is defined in \cite{channel}.
}

\subsection{ NOMA Network  Model}
To  improve the spectral efficiency of the system,  the NOMA transmission protocol is adopted in this paper, i.e., all users share the same frequency resource. 
Based on the above channel model, the incident signal of RIS    in (\ref{m1}) can be expressed as $\textbf{x}(t)=\textbf{H}_{B,S}(t)\sum_{k=1}^{K}p_kU_k(t)x_k(t)$, where $p_k$ is the constant transmit power, $x_k$ is unit power
information symbol of user $k$, $U_k$ is  a binary symbol to indicate whether user $k$ is communicating or not. Specifically, $U_k(t)=1$ indicates that user $k$ is communicating at slot $t$, otherwise, $U_k(t)=0$. 


As shown in Fig \ref{Fig1}, each communicating user $k$  receives the signal from the BS via direct and reflected wireless links. Then, the received signal can be denoted as follows
\begin{equation}
\begin{aligned}
y_k(t)=&h_{B,k}\sum_{j=1}^{K}p_jU_j(t)x_j(t)+\mathbf{H}_{S,k} \mathbf{y}_S+n_k
\\=&\left(\underbrace{h_{B,k}}_\text{direct link}+\underbrace{\mathbf{H}_{S,k}\mathbf{P\Theta}\mathbf{H}_{B,S}}_\text{RIS    reflected link}\right)\sum_{j=1}^{K}U_j(t)p_jx_j(t)+\underbrace{\mathbf{H}_{S,k}\mathbf{P\Theta z}}_{{\text{noise introduced \textcolor{black} {by} active RIS}}}+\underbrace{n_k}_{{\text{noise introduced \textcolor{black}{at} user $k$}}},
\end{aligned}
\end{equation}
where $n_k$ represents the additive white Gaussian
noise (AWGN) at user $k$ with zero mean and variance $\delta^2$. For ease of presentation,  the equivalent channel from
the BS to  the $k$-th user is defined as $h_k=h_{B,k}+\textbf{H}_{S,k}\mathbf{P\Theta}\textbf{H}_{B,S}$.  
Therefore, after transmitting through the RIS   decorated integrated channel, the signal received at user $k$ can be calculated as

\begin{equation}
\begin{aligned}
y_k(t)=\underbrace{U_k(t)h_k(t)p_kx_k(t)}_\text{desire signal}+\underbrace{\sum_{j\neq k}^{K}U_j(t)h_j(t)p_jx_j(t)}_\text{inte-user 	interference}+\underbrace{\mathbf{H}_{S,k}\mathbf{P\Theta z}+n_k}_\text{noise}.
\end{aligned}
\end{equation}

To eliminate the inter-user interference, SIC is carried out. \textcolor{black}{Note that the active RIS cannot perform channel estimation due to the lack of signal processing capability. In this paper, we assume that the system is performing channel estimation for the equivalent channel $h_k$, rather than estimating $\textbf{H}_{S,k}$, $\textbf{H}_{B,S}$ and $h_{B,k}$ separately. Moreover, to be more practical, imperfect SICs due to imperfect CSI is considered in the decoding process of NOMA. }  Specifically, each user determines decoding order based on the signal strength, which depends on the transmitted power and channel gains of the  users. The user with the strongest signal strength will be decoded first. Based on the principle of SIC,  each decoded user's signal will be regenerated and then subtracted from the remained signal. The signals of those users who failed to be decoded and also those who have not been decoded will be all regarded as  interference \cite{sic_faiL_next}.  For the  decoded user $k$ with a received signal strength of $g_k=U_k(t)|h_k(t)|^2p_k(t)$, it will be subjected to interference  as
\begin{equation}
\begin{aligned}
I_k(t)=\sum_{j\in{K},j\neq k}U_j(t)\left[1-\chi_{jk}(t)d_j(t)\textcolor{black}{\xi}\right]|h_j(t)|^2p_j(t),
\end{aligned}
\end{equation}
where $\chi_{jk}$ is a binary indicator with  $\chi_{jk}=1$  if the signal strength of the $j$-th user is stronger than the currently decoded user $k$, i.e., $|h_j(t)|^2p_j(t)>|h_k(t)|^2p_k(t)$, and $\chi_{jk}=0$  otherwise. Besides, we use $d_j(t)=1$ to indicate that the $j$-th user has been successfully decoded, and $d_j(t)=0$  to indicate that it has failed to be decoded or has not been decoded yet. \textcolor{black}{The parameter $0 < \xi < 1$ is used to characterize the  decoding error  due to imperfect CSI and hardware limitation. A larger value of $\xi$ indicates a smaller SIC error \cite{csi,ziwcl}.
}
Then, the achievable communication rate of user $k$ can be given by
\begin{equation}
\begin{aligned}
R_k(t)=U_k(t)\text{log}_2\left(1+\dfrac{g_k(t)}{I_k(t)+\parallel\mathbf{H}_{S,k}(t)\textcolor{black}{P}\mathbf{\Theta}(t)\parallel^2\delta_z^2+\delta^2}\right).
\label{r1}
\end{aligned}
\end{equation}
For successful decoding, it must satisfy the QoS requirements of users as 
$R_k(t)\geq R_0$,
where $R_0$ is the  rate threshold of each user.

\subsection{EH model}
\textcolor{black}{Due to the introduction of active components, the active RIS  consumes additional power to amplify the reflected signal. In view of the green energy-saving concept of RIS, EH technology is used to supply energy to active RIS. Considering that RIS is generally installed on the surface of high-rise buildings, where there are geographical advantages for solar and RF energy harvesting. Therefore, this paper adopts a hybrid energy harvesting framework that can overcome the dynamic climate problem in solar energy supply and the problem of collecting too little energy in RF energy supply \cite{inforcom}.}

\textcolor{black}{For the RF energy, we adopt a non-linear EH model based on the logistic function as in \cite{EH_nonlin}. The total harvested energy is modeled as
\begin{equation}
\begin{aligned}
	&E_{RF}(t)=\dfrac{\Psi(t)-E_M\Omega}{1-\Omega},  \quad \Omega=\dfrac{1}{1+exp(ab)},\\
&\Psi(t)=\dfrac{E_M}{1+exp(-a(P_{RF}-b))} ,
\end{aligned}
\end{equation} 
where $E_M$ is a constant denoting the maximum harvested RF power when the RF EH circuit is saturated.
The parameters $a$ and $b$ are constants related to the detailed circuit specifications. $P_{RF}=\rVert\textbf{H}_{B,S}\rVert^2$ denotes the receive RF power.}

\textcolor{black}{For the solar energy, we install a fully charged solar panel  with an area of $S_{sol}$ on the surface or on top of the building where the RIS is installed. We estimate harvested solar energy  by  the empirical model presented in \cite{inforcom}.  The EH model provides a year-round analysis of solar radiations and relates power levels to a quadratic equation on the time $t$ of the day,
\begin{equation}
	E_{sol}=S_{sol}(a1(t+a2)^2+a3)(1-\sigma_{sol}),
\end{equation}
where the parameters $a1$, $a2$, and $a3$ are vary seasonally for different months. $\sigma_{sol}$ is the percentage of cloud cover from weather reports. The  harvested RF and solar energy $e^h=E_{RF}+E_{sol}$  will be  stored in a rechargeable battery, which is available for signal amplification and reflection at the beginning of  next time slot.  Let the residual energy  stored in the battery  at the beginning of time slot $t$ be $E(t)$.  }

%
   
The energy stored in the battery will be scheduled for the amplification and transmission of the incident signal. At time slot $t$, the energy consumed by the active RIS  is
\begin{equation}
e^c(t)=\parallel \mathbf{P}\mathbf{\Theta}\mathbf{H}_{B,S}\parallel^2+\parallel\mathbf{P\Theta}\parallel^2\delta_z^2.
\label{ec}
\end{equation}
To ensure that the RIS    can successfully amplify the incident signal and transmit the amplified signal, it needs to satisfy $0\leq e^c(t)\leq E(t)$. Then, the $E(t)$ can be evolved as

\begin{equation}
\begin{aligned}
E(t+1)=
\min\{E(t)+e^h(t)-e^c(t), E_{max}\},
\label{B}
\end{aligned}	
\end{equation}
where $E_{max}$ denotes the maximum capacity of the battery.

\subsection{Optimization Problem Formulation}

Due to the stochastic nature of the UCS in the  RIS-aided EH-NOMA system and the dynamic nature of the harvested energy at the RIS, a rational control of the RIS  is necessary to improve the system performance. Specifically, the amplification  matrix $\textbf{P}$ and the phase shift matrix $\mathbf{\Theta}$ of the active RIS    are designed to maximize the successful communication  ratio, which is defined as  $\dfrac{\sum_{k=1}^K\mathbb{1}(R_k(t)\geq R_0)}{\sum_{k=1}^{K}U_k(t)}$. The $\mathbb{1}(\cdotp)$ is the indicator function, when the condition $(\cdotp)$ is satisfied, it is equal to 1, otherwise it is 0. Then, the optimization problem can be formulated as follows

\begin{equation}
\nonumber
\begin{aligned}
	\setlength{\abovedisplayskip}{3pt}
\setlength{\belowdisplayskip}{3pt}
\textbf{(P1):}\quad&\max_{\mathbf{P,\Theta}}\dfrac{\sum_{k=1}^K\mathbb{1}(R_k(t)\geq R_0)}{\sum_{k=1}^{K}U_k(t)}\\
\text{s.t.  }&U_k(t)\in\{0, 1\};& (C1) \\
&R_k(t)\geq R_0 ;&(C2)\\
&0\leq e^c(t)\leq E(t);& (C3)\\
&E(t+1)=
\min\{E(t)+e^h(t)-e^c(t), E_{max}\};&(C4)\\
&0\leq p_m^S\leq L,\quad \theta_m\in[0,2\pi) ,&(C5)
\end{aligned}
\label{opt}
\end{equation}
where $(C2)$  denotes to the QoS requirement for each user.   $(C3)$ defines and limits the energy consumed by the RIS which can be supplied by the stored energy. $(C4)$  corresponds to the energy  extrapolation principles.  $(C5)$  defines the constraints on the amplification factor and phase shift in the active RIS.

\section{LSTM based  Prediction Algorithm}

We propose an LSTM-DDPG algorithm to solve the optimization problem $(\mathbf{P1})$ for RIS controlling. As shown in Fig. \ref{Fig2}, the LSTM-DDPG algorithm is composed of the LSTM network and the DDPG framework. The LSTM network is used to extract the active 	RIS-aided EH-NOMA network  features and the DDPG framework is adopted to make controlling decisions. 

In this section, we first introduce a dynamic UCS model, which is quite different from the assumption that users are always in a communication state in most of the existing literature. Then an LSTM based algorithm  is designed for UCS prediction.

\subsection{Dynamic Communication State}

In most existing related works,   the users are usually assumed to be in a state of communication all the time, i.e., the users always have data to transmit. Such absolute assumption greatly reduces the dynamics and complexity of the system, thus reducing the difficulty of the associated network research. However, in practical wireless communication systems, such as IoT, vehicular networks, wireless sensor networks, and mobile communication networks, users will not always have data to transmit. 
Therefore, in this paper, the UCS are assumed to be dynamic. In each time slot, different users communicate based on different probabilities. And the probabilistic data of each user is assumed to be generated based on an independent random walk model \cite{LSTM2}.


In general, the UCS information can be obtained by analyzing real-time pilot signals in wireless networks. However, due to hardware limitations, the active RIS cannot handle the pilot signal \cite{active_ris,active_ris1}. Therefore, it cannot be informed of the dynamic UCS, nor can it be informed of the channel information of both BS-RIS and RIS-users links through the pilot signals. 
However, the channel information or the UCS are critical to RIS  control \cite{huang}.

\textbf{Remark 2.} \textit{A large number of RIS elements makes it too difficult to estimate the BS-RIS and RIS-users channels. And considering that the positions of the BS, RIS, and users are fixed, the fluctuation of channel information is smaller than the fluctuation of user communication states. Therefore,  to  control the active RIS,  LSTM is adopted to predict the UCS.}

LSTM is an improved version of recurrent neural networks (RNN). Unlike RNN which can only consider some recent states, by introducing the forget gates and input gates, LSTM is able to remember useful states in the long term and can also choose to forget some insignificant states.  Therefore, LSTM is often preferred when dealing with long-term time-dependent problems.

\subsection{LSTM based Prediction Algorithm for UCS}

\begin{figure}[t]
	\centering
	\includegraphics[width=6.5in]{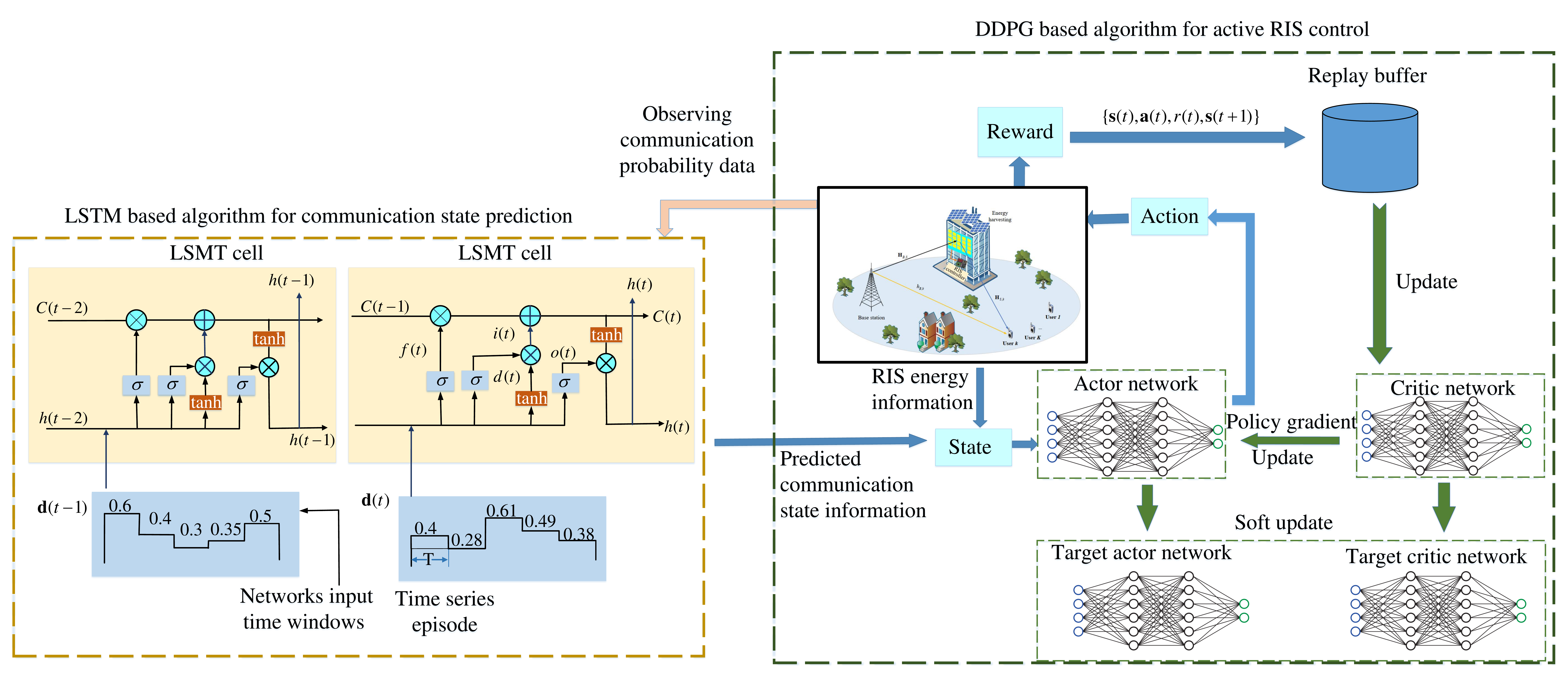}
	\caption{Structure diagram of RIS control algorithm based on LSTM-DDPG. }
	\centering
	\label{Fig2}	
\end{figure}

As an enhanced recursive network, LSTM can connect historical information to the current task. In this paper, since the communication probability of a user varies between each time slot, we express it as a time series when predicting the communication probability of each user. The time series data  are the input to LSTM along the  chain structure in a forward direction. We define the dynamic series of a user  as $\textbf{D}=[\textbf{ d}(1),\textbf{ d}(2),\cdots,\textbf{ d}(T_p)]$, where $\textbf{ d}(t)=[\text{Pr}(t-T_s),\text{Pr}(t-T_s+1),\cdots,
\text{Pr}(t-1)]$ are the time series of the input LSTM, and $Pr_k(t)$ is the communication probability  of one user at slot $t$.  $T_p$ is the total number of samples collected, and $T_s$ is the time interval. Note that since the prediction algorithm for the communication state is the same for each user, the user identifier $k$ is implicitly removed in this section for ease of presentation.

As shown in Fig. \ref{Fig2}, we use the classical LSTM structure \cite{LSTM2,LSTM4}. Each LSTM cell contains several hidden layers, which are forget gate, input gate, and output gate. 

\textbf{Forget gate:} 
The first layer is forget layer, which is also known as  $f(t)$. The main role of the forget gate is to decide which parts of the cell state $C(t-1)$ of the previous moment will be saved in the current state $C(t)$. It consists of the information passed from previous layer $h(t-1)$ and current input $\textbf{ d}(t)$ with weights $\hat{\textbf{w}}^f$, $\textbf{w}^f$, and bias $\textbf{b}^f$, which can be expressed as 
\begin{equation}
	{f}(t)=\sigma\left(\mathbf{w}^f(t)\mathbf{d}(t)+\mathbf{\hat{w}}^f(t)h(t-1)+\mathbf{b}^f(t)\right),
\end{equation}
where $\sigma$ represents the gate activation function which is normally sigmoid function. Due to the control of the forget gate, the LSTM can save information from a long time ago.

\textbf{Input gate:}
The input gate is used to prevent irrelevant content from entering the memory. It is used to determine how much of the input $\textbf{d}(t)$  at the current moment will be saved into the state $C(t)$ of the LSTM layer.
The input gate is achieved by a `sigmoid'  function  and a `tanh'    function, which can be  given as follow 
\begin{equation}
\begin{aligned}
	&{i}(t)=\sigma\left(\mathbf{w}^i\mathbf{d}(t)+\mathbf{\hat{w}}^i(t)h(t-1)+\mathbf{b}^i(t)\right),
\\
{d}&(t)=\text{tanh}
\left(\mathbf{w}_k^d \mathbf{d}(t)+\mathbf{\hat{w}}^d(t)h(t-1)+\mathbf{b}^d(t) \right).
\end{aligned}
	\end{equation}
\textbf{Output gate:} The output gate is used to control how much of the current  state  $C(t)$ is fed into the output $X(t)$. From Fig. \ref{LSTM}, we can see that ${C}(t-1)$, as the output state of the previous LSTM layer, will be fed  into the current LSTM layer, and the output state ${C}(t)$ of the current LSTM layer also passes the next LSTM layer. The output state at current time $t$ of the layer can be updated as
\begin{equation}
C(t)={f}(t)\otimes C(t-1)+{i}(t)\otimes{d}(t).
\end{equation}
The output  is also achieved by the sigmoid function and `tanh' function, which can be calculated as
\begin{equation}
\begin{aligned}
	{o}(t)=&\sigma
\left(\mathbf{w}^o(t)\mathbf{d}(t)+\mathbf{\hat{w}}^o(t)h(t-1)+\mathbf{b}^o(t)\right),\\
&h(t)={o}(t)\otimes \text{tanh}(C(t)).
\end{aligned}
\end{equation}
With the network output $h(t)$, the estimate communication probability can be obtained as
\begin{equation}
	\hat{{Pr}}(t)=\mathbf{w}'(t){o}(t)\otimes \text{tanh}(C(t)),
\end{equation} 
where $\textbf{w}'$ is the regression coefficient. The  $\textbf{w}'$ and the above  parameters can be obtained by the real-time recurrent learning algorithm.

Based on the above three types of gate hidden layers, the LSTM  can predict the time series of users' communication probabilities based on the real communication information and the previously estimated communication probabilities. 
By assuming that the  estimate output of the   LSTM at moment $t$ is ${ \hat{\text{Pr}(t)}}$, and the observed true communication probability at that moment is ${\text{Pr}}(t)$, then the loss function can be calculated as 
\begin{equation}
	\text{Loss}(\bm{\theta}_l)=\dfrac{1}{T_p}\sum_{t=1}^{T_p}\left({\text{Pr}}(t)-\hat{\text{Pr}(t)}\right)^2,
\label{losses}
\end{equation}
where  $\bm{\theta}_l$ denotes  network parameters of the LSTM network.  Note that it is assumed that the real communication probability data used for training is available through observation, and such an assumption is often used in  probabilistic data prediction algorithms \cite{LSTM2}.
\begin{spacing}{1.1}
	\begin{algorithm}[!t] 
		\caption{The training and application process of the LSTM based   prediction algorithm} 
		\label{ag11} 
		\begin{algorithmic}[1] 
			\STATE{\textbf{Training Process:}}	
			\REQUIRE  
			Training data: 70\% of the true communication probability sequence  $\textbf{D}$. 
			\ENSURE 
			Trained LSTM model, i.e., the optimal network parameters  $\bm{\theta}^*_l$.
			\STATE{\textbf{Initialization:} Initialize the networks parameters $\bm{\theta}_l$. }
			\FOR{each episode}
			\STATE{Choose the network input $s(t)$ from the observed training  data;}
			\STATE{Obtain the output of the last LSTM layer $h(t)$, and calculate the estimate output $\hat{\text{Pr}(t)}$ ;}
			\STATE{Calculate the loss function $\text{Loss}(\bm{\theta}_l)=\dfrac{1}{T_p}\sum_{t=1}^{T_p}\left({\text{Pr}}(t)-\hat{\text{Pr}(t)}\right)^2$; }
			\STATE{Update network parameters $\bm{\theta}_l$ by minimizing the loss function.}
			\ENDFOR
			\STATE{\textbf{Application Process:}} 
			\REQUIRE  
			Probability of  communication for the previous 5 time slots.	\ENSURE 
			The predicted communication state of the user at the current moment.
			\FOR {each step $t$}
			\STATE {Input the previous 5 time slot communication probability data to the LSTM networks;}
			\STATE{The LSTM output the estimated communication probability $\hat{\text{Pr}}(t)$;}
			\STATE{Obtain the UCS $\hat{U}(t)$ based on the estimated probability.}
			\ENDFOR
		\end{algorithmic}
	\end{algorithm}
\end{spacing}
The Algorithm \ref{ag11} specifies the training and application process of the proposed LSTM based prediction algorithm, which is trained and  applied in the RIS controlled.  
After training convergence, this LSTM based prediction algorithm will be applied to the prediction of UCS.  The algorithm remedies the hardware deficiency of the active RIS.  \textcolor{black}{Note that in the LSTM-based UCS prediction algorithm, the input to the algorithm is continuously updated historical data. Specifically, the information related to the UCS at the current moment is transformed into the historical data at the next moment \cite{LSTM2,LSTM4}.}


\section{LSTM-DDPG based Algorithm for Active RIS Control}


In this section,  we first model  optimization problem  as a Markov decision processes (MDPs). Then a DDPG algorithm is designed for controlling the amplification  and phase shift matrix.
\subsection{ MDPs Formulation}
 The optimization problem $(\textbf{P1})$ can be formulated as an  MDPs, which  consists of an agent, a set of environment sate $\textbf{s}(t)$,  a reward function $R(t)$,  and a set of action  $\textbf{a}(t)$. 

\textbf{Agent:}
Considering that the optimization objects of problem  $(\textbf{P1})$ are the amplification matrix  $\textbf{P}$ and the phase shift matrix  $\bm{\Theta}$ of RIS,   it is natural to choose the  RIS  controller  as the only agent for the system.

\textbf{Environment state:}
Due to the simple structure of the active RIS, it does not have the ability to analyze and process the pilot  signal. Therefore, the agent  is unable to obtain the channel information of BS-RIS     and RIS-user links. Stepping back, we use the predicted communication state  $\hat{U}_k(t),\forall k\in {K}$ of all users as part of the  environment state. 

Besides, since the active RIS    is powered by the EH, i.e., the available energy   $E(t)$ is dynamic, which  has a significant impact on the decision making of the agent.  Therefore, the currently available energy value $E(t)$  at the RIS    is designed as part of the state. In summary, the environmental state of the RIS-aided EH-NOMA system    at slot $t$ can be expressed as
\begin{equation}
	\setlength{\abovedisplayskip}{3pt}
\setlength{\belowdisplayskip}{3pt}
\textbf{s}(t)=\left[\hat{U}_1(t),\hat{U}_2(t),\cdots,\hat{U}_K(t), E(t)\right].
\label{state}
\end{equation}
Clearly, at the beginning of each time slot, the predicted communication state $\hat{U}_k(t), \forall k\in {K}$,   and remaining battery energy $E(t)$ can be locally observed by the agent. 

%

\textbf{Action:}
The action  is composed of two matrices, the  amplification matrix $\textbf{P}$ and the phase shift matrix $\bm{\Theta}$, which  is naturally defined as
\begin{equation}
	\mathbf{a}(t)=\left[p_1^S,p_2^S,\cdots, p_M^S; \theta_1,\theta_2,\cdots,\theta_M\right].
	\label{action}
\end{equation}

The  element $p_m^S$ in $\textbf{P}$ and the element $\theta_m$ in $\bm{\Theta}$ characterize the power amplification and phase of the $m$-th RIS    element, respectively. And they satisfy
\begin{equation}
	0\leq p_m^S\leq L, \quad 0\leq\theta_m< 2\pi, \quad \forall m\in {M}.
\end{equation}
In addition, since the  RIS    is powered by the EH, its available energy is dynamic and limited. This will lead to the fact that the available energy may not be able to support some of the actions taken by the agent in the early stages of training.  Therefore, in the training of the DRL, when the agent makes an action decision at each time slot, the RIS  first checks whether its available energy is sufficient to perform the action. If it is not enough, the action needs to be adjusted as follows
\begin{equation}
\setlength{\abovedisplayskip}{1pt}
\setlength{\belowdisplayskip}{1pt}
\left[\bar{p_m^S}(t),\quad \bar{\theta_m}(t)\right] =
\begin{cases}
\left[p_m^S(t),\quad \theta_m(t)\right]&   \quad e^c(t)\leq E(t) \\
\left[\dfrac{1}{L}p_m^S(t),\quad \theta_m(t)\right]&   \quad {e^c_0}(t)\leq E(t)<e^c(t) \\
[0,\quad 0]&   \quad e^c_0(t)> E(t)
\end{cases},
\label{aa}
\end{equation}
where $e^c_0(t)=|| \mathbf{\Theta}\mathbf{H}_{B,S}||^2+||\mathbf{\Theta}||^2\sigma_z^2$ is the energy consumed by RIS    without power amplification (i.e. $0\leq p_m^S\leq1$). Obviously, the design and adjustment strategy of the action vector can guarantee the constraints on energy harvesting, power amplification, and phase shift in the optimization problem $(\textbf{P1})$.

\textbf{Reward function:}
In the DRL algorithm, the reward function plays a crucial role in motivating agent to find optimal policy faster. Considering that the objective of the optimization problem $(\textbf{P1)}$ is to maximize the successful communication ratio, a brief reward function is designed as follows
\begin{equation}
	\setlength{\abovedisplayskip}{3pt}
\setlength{\belowdisplayskip}{3pt}
R(t)=r(\sum_{k=1}^{K}\mathbb{1}_{R_k(t)\geqq R_0}),
\label{reward_all}
\end{equation}
where $r$  is  a positive multiplier greater than 1 that gives the agent an appropriately larger reward to speed up the algorithm training.  Clearly,  to get higher rewards, the agent will try to learn a better policy to achieve a higher successful communication ratio.


\subsection{LSTM-DDPG based RIS  Control Algorithm}
\subsubsection{DDPG }
For the formulated MDPs framework, as indicated by (\ref{state}) and (\ref{action}), the state and action space are both continuous and multi-dimensional. Therefore, compared to value based DRL algorithms, such as Q learning and DQN  algorithms \cite{dqn}, the  policy gradient DRL algorithms, such as AC and DDPG are more suitable for the optimization problem in this paper. The structures of the DDPG and AC algorithm are similar, both consisting of an actor network and a critic network, except that DDPG has two more target networks, i.e., the target actor network and the critic network. In fact, DDPG is an upgraded algorithm of AC, which combines the advantages of DQN and AC. Therefore, the DDPG algorithm is adopted in this paper for the design of active RIS.
As shown in Fig. \ref{LSTM},  four DNNs are included in the DDPG  algorithm to  learn the optimization  policy for the active RIS, which are listed as follows:

\textbf{Actor network:} The actor network is also known as the policy network with  parameters $\bm{\theta}_{\mu}$. It is dedicated to learning the decision parameterized  policy ${\mu}$ of the whole algorithm. Based on the learned policy, it  outputs the corresponding action $\textbf{a}(t)=\mu\left(\textbf{s}(t)\mid\bm{\theta}_{\mu}\right )$ for any environmental state $\textbf{s}(t)$.

\textbf{Critic network:} The critic network is also known as the Q network with  parameters $\bm{\theta}_{Q}$.  The main role of the critic network is to evaluate the policy $\mu$ learned by the actor network and thus guide the direction of parameter updates for the actor network. It takes the current environment state $\textbf{s}(t)$ and action $\textbf{a}(t)$ as the network input, and outputs the corresponding  state-action
value $Q(\textbf{s}(t),\textbf{a}(t)\mid\bm{\theta}_Q)$.

\textbf{Target  networks:} The target actor network with  parameter $\hat{\bm{\theta}_{\mu}}$  and the target critic network with  parameter $\hat{\bm{\theta}_{Q}}$  are introduced to calculate the target action and the target Q value $Q'$, respectively. They can effectively improve the stability and convergence of the DDPG algorithm. The network structures of the target actor (critic) network and the main actor (critic) network are exactly the same. The difference between the two types of networks is that there is a time interval on the update of the network parameters. 
Specifically, the target network parameters are updated  through soft updating every $C$ steps as follows
\begin{equation}
\begin{aligned}
\bm{\theta}_{\mu'}(t)&=\tau_\mu \bm{\theta}_{\mu }(t)+(1-\tau_\mu)\bm{\theta}_{\mu'}(t),\\
\bm{\theta}_{Q'}(t)&=\tau_Q \bm{\theta}_{Q}(t)+(1-\tau_Q)\bm{\theta}_{Q'}(t),
\end{aligned}
\label{soft}
\end{equation}
where $0<\tau_\mu\ll1$ and $0<\tau_Q\ll1$ denote the soft updating factors.  

\subsubsection{The training process of LSTM-DDPG based control Algorithm}
In each time slot $t$, 
the agent observes the active RIS    aided EH-NOMA system  and obtains an environment state $\textbf{s}(t)$.  
The state is then fed into the actor network,  which outputs the corresponding action $\textbf{a}(t)$. In order  to fully explore the environment, the output actions need to be noised  during the training, i.e.
\begin{equation}
\mathbf{a}(t)=\mu\left(\mathbf{s}(t)\mid\bm{\theta}_\mu(t)\right)+n_{o}(t),
\end{equation}
where $n_o(t)$ is the exploration noise, which can be expresses as 
\begin{equation}
n_o(t)=
\begin{cases}
n_{ini}-t*\varphi, &
n_o(t)>n_{end}\quad \\
n_{end},& \text{otherwise}
\end{cases},
\label{anoise}
\end{equation}
where $n_{ini}$, $n_{end}$, and $\varphi$ represent the maximum exploration noise, the minimum exploration noise, and the decreasing factor of the exploration noise, respectively. Note that the action output by the actor network are first noise-added based on (\ref{anoise}), and then the action are adjusted based on (\ref{aa}). Then, the agent obtains the corresponding reward $R(t)$, and the system environment moves to the next state $\textbf{s}(t+1)$.
Based on this exploration, the agent can obtain an experience tuple $\left[\textbf{s}(t),\bar{\textbf{a}}(t),R(t),\textbf{s}(t+1) \right]$, which will be stored in the experience replay memory $\mathcal{D}$ and used for the training of the neural network. The main idea of training is to obtain an optimal RIS  control policy $\mu*$ , which can be realized by updating the parameters of  actor and critic networks until  convergence. 

\begin{spacing}{1.0}
	\begin{algorithm}[!t] 
		\caption{Training Process of LSTM-DDPG  Algorithm} 
		\label{alg}
		\linespread{2}
		\begin{algorithmic}[1] 
			\REQUIRE  
			The LSTM networks with  network parameters  $\bm{\theta}^*_l$, the learning rate  $\beta_\mu$ and $\beta_Q$, the soft updating factor $\tau_\mu$ and $\tau_Q$.
			\ENSURE 
			The trained actor and critic networks with optimal network parameters $\bm{\theta}_{\mu}^*$ and  $\bm{\theta}_{Q}^*$.
			\STATE \textbf{Initialize :} Empty the replay memory $\mathcal{D}$, $\bm{\theta}_{\mu}$ and $\bm{\theta}_Q$ with random weighs and bias.
			\FOR{each episode $T=1,\ldots,T^{all}$}
			\STATE{Reset the RIS-aided EH-NOMA system;}
			\FOR {each step $N=1,\ldots,N^{all}$}
			\STATE{Predict the UCS by the trained LSTM network; }
			\STATE{Observe the  state $\textbf{s}(t)$ based on (\ref{state}) and input it into the actor network;}
			\STATE{The actor network outputs the corresponding action $\textbf{a}(t)$	and adjusts it based on (\ref{aa});}
			\STATE{Adding noise to the adjusted action by (\ref{anoise}), and then execute it on the network};
			\STATE{The agent calculates the local reward $R(t)$ according to (\ref{reward_all}) and the environment enter to the next state $\textbf{s}(t+1)$};
			\STATE{Stores the experience tuple $\left({\textbf{s}(t),{\textbf{a}}(t)(t),R(t),\textbf{s}(t+1)}\right)$ into Memory $\mathcal{D}$};
			\STATE {Soft update the target networks of central trainer according to (\ref{soft})};
			\IF{No. stored tuples $ \geq\dfrac{1}{3}|\mathcal{D}|$}
			\STATE{Sample a mini-batch with $\Omega$ transitions from $\mathcal{D}$};
			\STATE{Update the critic network by minimizing the loss in (\ref{lossQ})};
			\STATE{Update the actor network by maximizing the policy 	gradient in (\ref{gra})}.
			\ENDIF
			\ENDFOR	
			\ENDFOR	
		\end{algorithmic}
	\end{algorithm}
\end{spacing}
To updating the actor and critic networks,   the $\Omega$-size mini-batch   will be randomly sampled from $\mathcal{D}$. The update of critic network parameters is achieved by reactive transfer of TD errors, and the  loss function is defined as
\begin{equation}
\setlength{\abovedisplayskip}{3pt}
\setlength{\belowdisplayskip}{3pt}
L(\bm{\theta}_Q)=\dfrac{1}{\Omega}\sum_{i=1}^{\Omega}
\left[Q'(i)-Q\left(\mathbf{s}(i),\mu(\mathbf{s}(i)|\bm{\theta}_\mu)|\bm{\theta}_Q(i)\right)\right]^2,
\label{lossQ}
\end{equation}
where $Q'$ is the target value of the state-value function (i.e., the output of target critic network), which can be calculated by the Bellman equation  as
\begin{equation}
	\setlength{\abovedisplayskip}{3pt}
\setlength{\belowdisplayskip}{3pt}
Q'(i)=R(i)+\gamma \mathop{\max}_\textbf{a} Q'\left(\mathbf{s}(i+1),\mu'(\mathbf{s}(i+1)|\bm{\theta}{\mu' }(i))|\bm{\theta}_{Q'}(i)\right).
\end{equation}
Then by minimizing the loss function (\ref{lossQ}), the critic network parameters can be updated as follows
\begin{equation}
	\setlength{\abovedisplayskip}{3pt}
\setlength{\belowdisplayskip}{3pt}
\bm{\theta}_Q\leftarrow \bm{\theta}_Q-\beta_Q\triangledown_{\bm{\theta}_Q}L(\bm{\theta}_Q),
\end{equation}
where $\beta_Q$ is the learning rate of the critic network.

The optimization goal of the actor network is to obtain the maximum state-action function Q.  Hence, considering the fact that the state-action function Q is differentiable and the action space  is continuous,  the actor network can be updated by the policy gradient with the ascent factor as follows
\begin{equation}
\begin{aligned}
	\setlength{\abovedisplayskip}{3pt}
\setlength{\belowdisplayskip}{3pt}
\triangledown_{\bm{\theta}_\mu}J(\bm{\theta}_\mu)  =\dfrac{1}{\Omega}\sum_{i=1}^{\Omega}\triangledown_\mathbf{a}Q(\mathbf{s},\mathbf{a}|\bm{\theta}_Q)|_{\mathbf{s}=\mathbf{s}(i),\mathbf{a}=\mu(\mathbf{s}(i))}\triangledown_{\bm{\theta_\mu}}\mu(\mathbf{s}|\bm{\theta}_\mu)|_{\mathbf{s}(i)}.
\end{aligned}
\label{gra}
\end{equation}
Algorithm \ref{alg} gives the specific training  process of the LSTM-DDPG based RIS control algorithm, which is embedded in the RIS controller.


Through sufficient exploration and training, DDPG will converge and learn the optimal policy $\mu^*(\textbf{s}|\bm{\theta}_{\mu}^*)$ for controlling the RIS   amplification power matrix and phase shift matrix. 
If there are no large fluctuations in the system parameters (e.g., $M$, $K$ $E^H_{max}$), the network in DDPG no longer needs to be retrained.  In the online applications, based on the learned policy $\mu^*$, the active RIS   can   design  $\textbf{P}$ and $\bm{\Theta}$ reasonably for the current system environment. 


\subsection{Complexity Analysis and Convergence Proof }
\subsubsection{Complexity Analysis of LSTM  algorithm}
The LSTM based algorithm contains a three-layer network structure, i.e., an input layer,  LSTM layer, and an output LSTM layer. The activation function of both LSTM layer and output LSTM layer are `tanh', and the optimization function of the whole network is `RMSprop'. The number of neurons contained in the input layer,  LSTM layers, and output layer are defined as $L_i$, $L_1$, and $L_o$, respectively.  The input data size of the input layer is $L_i=6$, where the first 5 are the user communication probabilities for 5 time slots and the last one characterizes the dimension of the input data. The output of the LSTM algorithm is the communication probability of a user at the current moment, and its dimension is  $L_o=1$.  According to \cite{LSTM4}, the computational
complexity of the LSTM  layer  is $\mathcal{O}(W1)$, where $W1=4L_1(L_i+b1+L_1)$, and the complexity of the output LSTM layer is $\mathcal{O}(W2)$, where $\mathcal{O}(W2)=4L_2(L_1+b2+L_o)$, $b1$ and $b2$ are the bias in the two LSTM layers.
Then, the  complexity of the proposed LSTM based prediction algorithm is 
  $\mathcal{O}(W1+W2)$. 
\subsubsection{Complexity Analysis of DDPG  algorithm}
In the proposed DDPG based RIS control algorithm, there are four DNNs, i.e., actor and critic main networks, and two target networks with the same structure as the main networks. Among them, the actor network contains input layer, output layer and three hidden layers with $l_i$ neurons in the $i$-th layer. According to  (\ref{state}) and (\ref{action}), the number of neurons in the input layer and output layer of the actor network is $l_0^a=k+1$,   $l_4^a=2M$, respectively.  The rectified linear
activation function (ReLU) is used in both hidden layers,
and the sigmoid activation function is used for the output layer.  The actor network also contains an input layer, an output layer, and three hidden layers with $l_i^c$ neurons  in the $i$-th layer.

\begin{itemize}
\item Application complexity:
During the online application, only the actor network needs to be executed, and for any input environment state $\textbf{s}$, the trained actor network outputs  corresponding action $\textbf{a}$. According to
the connection and calculation principle of the actor network,
we can get a moderate computational complexity of the process from
input to output that is as  ${\mathcal{O}}(\sum_{i=0}^{3}l_i^{
	a}l_{i+1}^{
	a})$. 
\item Training complexity: In the training process, both the actor network and critic network need to be trained, and the most intuitive complexity is caused by the back propagation. Besides,  the training process needs the prediction
results from the target actor network and target critic network. Thus, a single back propagation training
step for the proposed DDPG structure will contribute the
complexity of 
$\mathcal{O}(\sum_{i=0}^{3}2l_i^{
	a}l_{i+1}^{
	a}+\sum_{i=0}^{3}2l_i^{
	c}l_{i+1}^{
	c})$.
In addition, it can be seen from Algorithm \ref{alg} that until the number of tuples stored in the replay memory $\mathcal{D} $ exceeds  $ \geq\dfrac{1}{3}|\mathcal{D}|$, the agent is only explored and not trained.
Therefore, for the whole training process, the overall complexity of the DDPG based algorithm can be calculated as $\mathcal{O}\left((N^{all}T^{all}-\dfrac{1}{3}|\mathcal{D}|)\Omega(\sum_{i=0}^{3}2l_i^{
	a}l_{i+1}^{
	a}+\sum_{i=0}^{3}2l_i^{
	c}l_{i+1}^{
	c})+\dfrac{1}{3}|\mathcal{D}|(\sum_{i=0}^{3}l_i^{
	a}l_{i+1}^{
	a})\right)$.

\end{itemize}

In this paper, the number of neurons in the hidden layer of the actor network and the critic network are set to $l_1^a=l_1^c=256$, $l_2^a=l_2^c=256$, and $l_3^a=l_3^c=128$. In this LSTM based algorithm, the numbers of neurons of two LSTM layers are set to $L_1=64$. Due to the simple structure of the LSTM network and the small numbers of neurons in each layer, the complexity of this LSTM algorithm is very low. 
The DDPG algorithm has higher complexity compared to LSTM, but its algorithm complexity is very low compared to the DRL algorithm with multiple agents. In addition, the  state of the DDPG is a local observation with low dimensionality and does not require additional interaction with the users, which also greatly reduces the application complexity of the LSTM-DDPG  algorithm. Based on the development and application of integrated development technology and software-defined networking technology, embedding LSTM  and DDPG  into the controller of RIS can be easily implemented with acceptable complexity.
\subsubsection{\textcolor{black}{Convergence Proof}}

\textcolor{black}{\textit{Theorem 1}: The  proposed DDPG algorithm can converges.
}

\textcolor{black}{\textit{Proof}: Since the state space and the action space  in the RIS-aided EH-NOMA wireless environment are finite, and all  pairs $\{(\mathbf{s(t)},\mathbf{a}(t)|t\in N^+) \}$ can be visited infinitely. 	In each training step $t$, the agent RIS can obtain an  experience tuple $\left({\textbf{s}(t),{\textbf{a}}(t)(t),R(t),\textbf{s}(t+1)}\right)$,then  the Q function of critic network can be  updated by:
	\begin{equation}
	\begin{aligned}
	Q(\textbf{s}(t),\textbf{a}(t))=&Q(\textbf{s}(t),\textbf{a}(t))+\alpha(\textbf{s}(t),\textbf{a}(t)\\
	&[R(t)+\gamma \mathop{\max}_\textbf{a} Q(\textbf{s}(t+1),\textbf{a})-Q(\textbf{s}(t),{\textbf{a}}(t))],
	\end{aligned}
	\end{equation}
	where $\alpha(\textbf{s},\textbf{a})=0$ unless $(\textbf{s},\textbf{a})=(\textbf{s}(t),\textbf{a}(t))$. The Q function is intended to approximate the optimal Q function $Q^*$ of the MDP. By subtracting $Q^*(\textbf{s}(t),\textbf{a}(t))$  from both the left and right sides of the above equation, the following equation can be obtained
	\begin{equation}
	\begin{aligned}
	\Psi(\textbf{s}(t),\textbf{a}(t))=Q(\textbf{s}(t),\textbf{a}(t))-Q^*(\textbf{s}(t),\textbf{a}(t)),
	\end{aligned}
	\end{equation}
	where
	\begin{equation}
	\begin{aligned}
	\Psi(\textbf{s}(t),\textbf{a}(t))=(1-\alpha)\Psi(\textbf{s}(t),\textbf{a}(t))+\alpha(\textbf{s}(t),\textbf{a}(t))G(\textbf{s}(t),\textbf{a}(t)),\\
	G(\textbf{s}(t),\textbf{a}(t))=[R(t)+\gamma\mathop{\max}_\textbf{a}Q(\mathbf{s}(t+1),\mathbf{a})-Q^*(\textbf{s}(t),\textbf{a}(t))].
	\end{aligned}
	\end{equation} 
	Because $\sum_{t=1}^{\infty}\alpha=\infty$  and $\sum_{t=1}^{\infty}\alpha^2<\infty$, according to \cite{conver},
	$\Psi(\textbf{s}(t),\textbf{a}(t))$ converges to zeros w.p.1 if:}

\textcolor{black}{1)$\| \mathbb{E}[G(\textbf{s}(t),\textbf{a}(t))|G]\|_{\infty}\leq \gamma \| \Psi(\textbf{s}(t),\textbf{a}(t))\|_{\infty}$ with $\gamma<1$.}

\textcolor{black}{2)$\text{var}[G(\textbf{s}(t),\textbf{a}(t))|G]\leq C(1+\| \Psi(\textbf{s}(t),\textbf{a}(t))\|_{\infty}^2)$, with a positive constant $C$.}

\textcolor{black}{First, the derivation of $\mathbb{E}[G(\textbf{s}(t),\textbf{a}(t))|G]\|_{\infty}$ is performed as follows
	\begin{equation}
	\begin{aligned}
	\mathbb{E}[G(\textbf{s}(t),\textbf{a}(t))|G]\|_{\infty}&=Pr(\mathbf{s}(t+1)|\mathbf{s}(t),\mathbf{a}(t))G(\textbf{s}(t),\textbf{a}(t))\\
	&\leq \gamma \| Q(\textbf{s}(t),\textbf{a}(t))-Q^*(\textbf{s}(t),\textbf{a}(t))\|_{\infty}\\
	&=\|\Psi(\textbf{s}(t),\textbf{a}(t))\|_{\infty}.
	\end{aligned}
	\end{equation}
	Then, the $\text{var}[G(\textbf{s}(t),\textbf{a}(t))|G]$ can be calculated as
	\begin{equation}
	\text{var}[G(\textbf{s}(t),\textbf{a}(t))|G]=\text{var}[R(t)+\gamma\mathop{\max}_\textbf{a}Q(\mathbf{s}(t+1),\mathbf{a})|G].
	\end{equation}
	Since
	$R(t)=r(\sum_{k=1}^{K}\mathbb{1}_{R_k(t)\geqq R_0})$ is bounded, then we can obtain
	\begin{equation}
	\text{var}[G(\textbf{s}(t),\textbf{a}(t))|G]\leq C(1+\| \Psi(\textbf{s}(t),\textbf{a}(t))\|_{\infty}^2).
	\end{equation} 
	Therefore, $\Psi(\textbf{s}(t),\textbf{a}(t))$ can converge to zero w.p.1, which means that the critic network of the proposed DDPG algorithm can converges to the optimal Q function $Q^*(\textbf{s},\textbf{a})$.}

\textcolor{black}{Then,  for any state $\textbf{s}$, the optimal action $\mathbf{a}^*$ can be selected  based on the optimal function $Q^*(\textbf{s},\textbf{a})$, i.e., 
	\begin{equation}
	\mathbf{a}^*(t)=arg \mathop{\max}_\textbf{a}Q^*(\mathbf{s}(t),\mathbf{a}).
	\end{equation}
	The proof is completed.
	$\hfill\blacksquare$}

\section{Simulation Results and Discussions }
In this section, we verify the performance of the proposed LSTM-DDPG  algorithm in downlink active RIS-aided EH-NOMA networks. First, the accuracy of the LSTM based UCS prediction algorithm is verified, and then we examine the performance advantages of NOMA over OMA in active RIS EH networks. In addition, we compare the performance of active and passive RIS in the EH-NOMA  network. Finally, the proposed LSTM-DDPG  algorithm is compared with several other benchmark algorithms.  Considering the actual communication scenario, the positions of both BS and RIS  are fixed, and their 3D position coordinates are set to (0 \text{m}, 0 \text{m}, 0 \text{m}) and (100 \text{m}, 100 \text{m}, 50 \text{m}), respectively. All users are on the right side of the BS and RIS, and they are randomly distributed in a semicircle with a radius of $200 \text{m}$  to $500 \text{m}$ from the BS.  \textcolor{black} {In all simulations, $\xi=0.9$ is set to characterize the NOMA decoding error due to imperfect CSI. The parameters related to solar and RF energy harvesting in the energy model are set with reference to \cite{inforcom} and \cite{EH_nonlin}, respectively. The energy of the fully charged solar panel is 50 joules.}  Unless otherwise specified,
the simulation parameters  are  given in Table \ref{table1}, which follows the simulation parameters setting in  \cite{ris_drl3,ris_drl2}. 

%
\begin{table}[!t]
	\small
	\caption{SIMULATION PARAMETER SETTING}
	\label{table1}
	\begin{center}	
		\renewcommand\arraystretch{2} 
		\begin{tabular}{|c|c|c||c|c|c|}			\hline
			\text{Parameter}&\text{Description} &\text{Value}& \text{Parameter}&\text{Description} &\text{Value}\\	 
			\hline	 
			$B$&bandwidth& 1 $ \text{MHz}$
			& $\tau_\mu,\tau_Q$& soft updating factor&$0.01$\\
			$r$&reward parameters&  10	
			&$p_k$ &transmitting power &0.1 \text{W}\\
			$E^H_{max}$& maximum arrive energy&0.6 \text{W}&$\eta$&EH efficiency coefficient&0.9	\\
			$| \mathcal{D}|$&capacity of memory  &10000&$C_0$ &pass loss at reference distance&-30 \text{dB}\\
			$T^{all}$& episode number& 200 &$N^{all}$&total step in each episode& 200    \\
			$\Omega$&size of mini-batch & 40
			& $\beta_\mu$, $\beta_Q$ &learning rate& $0.001$\\
			$\alpha_{B,k}$  & pass loss exponent BS-users&3.5	
			&$\alpha_{B,S}$,\quad $\alpha_{S,k}$	& pass loss exponent BS-RIS   &2.2\\
			\hline
		\end{tabular}
	\end{center}
\end{table}
\subsection{Performance Verification of LSTM based Prediction Algorithm}
\begin{figure}[t]
	\centering
	\subfigure[\textcolor{black}{user} 1]{\includegraphics[width=0.45\textwidth]{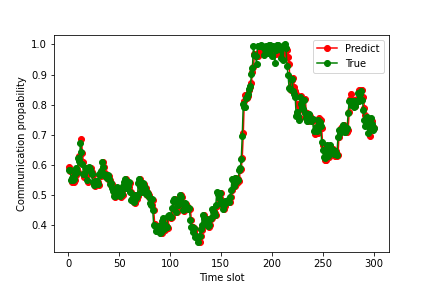}}
	\subfigure[\textcolor{black}{user} 2]{\includegraphics[width=0.45\textwidth]{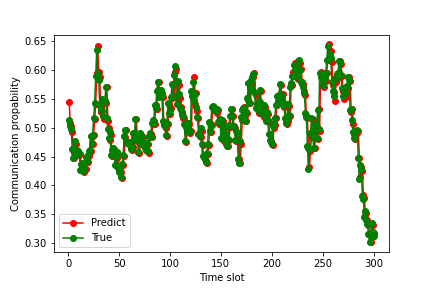}}
	\subfigure[\textcolor{black}{user} 3]{\includegraphics[width=0.45\textwidth]{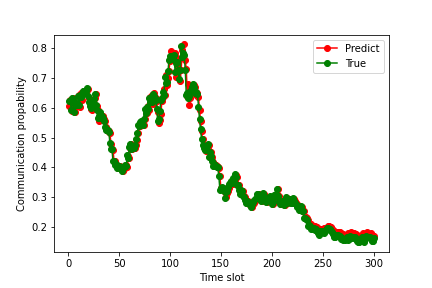}}
	\subfigure[\textcolor{black}{user} 4]{\includegraphics[width=0.45\textwidth]{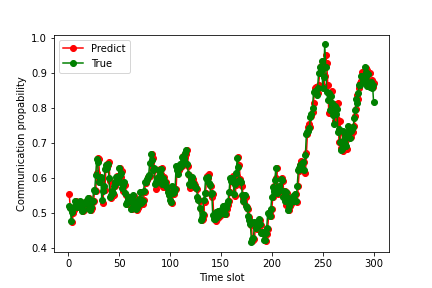}}
	\caption{Prediction effectiveness of LSTM based  algorithm.}
	\label{LSTM}
\end{figure}
\begin{figure}[t]
	\centering
	\subfigure{\includegraphics[width=0.49\textwidth]{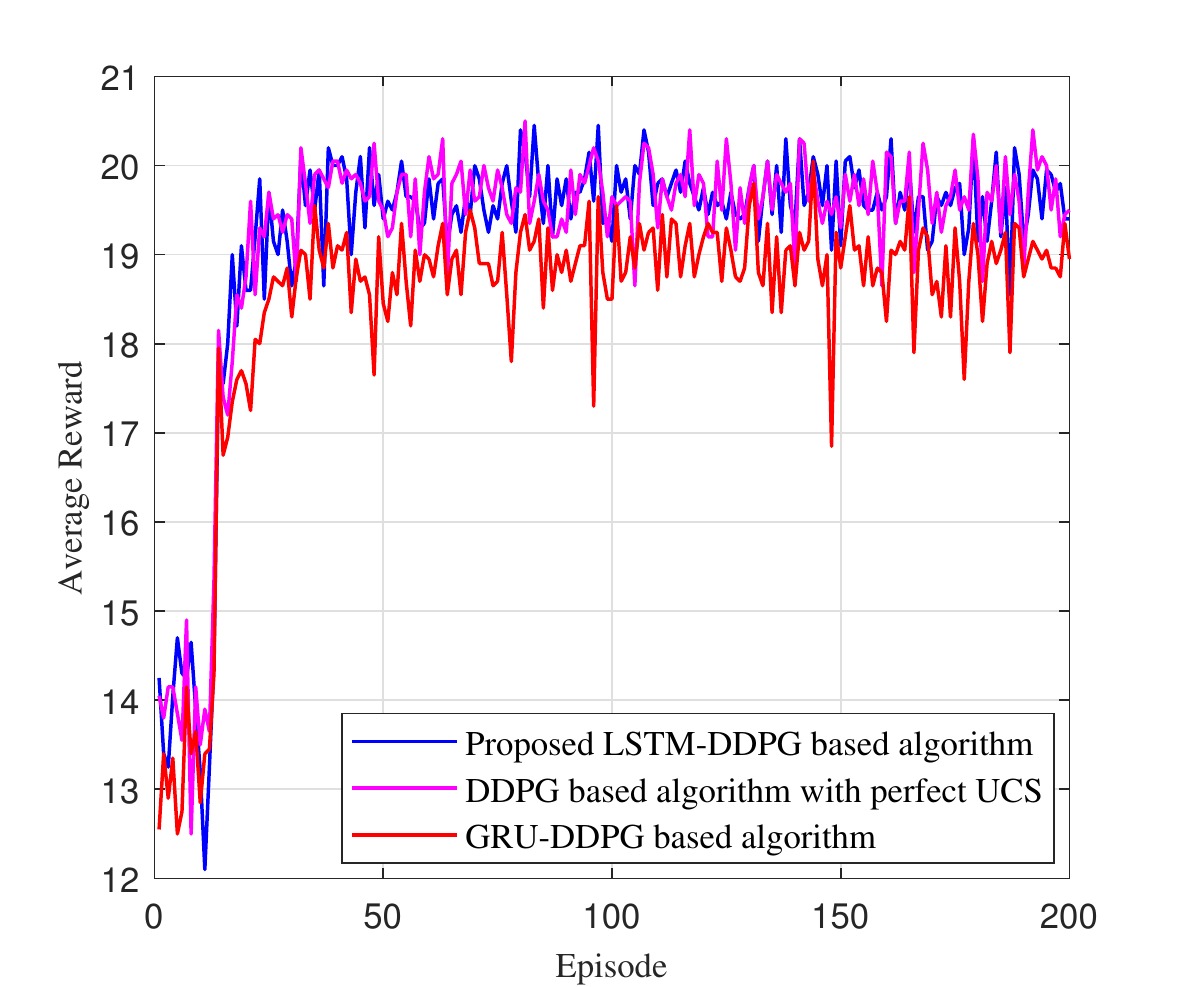}}
	\subfigure{\includegraphics[width=0.49\textwidth]{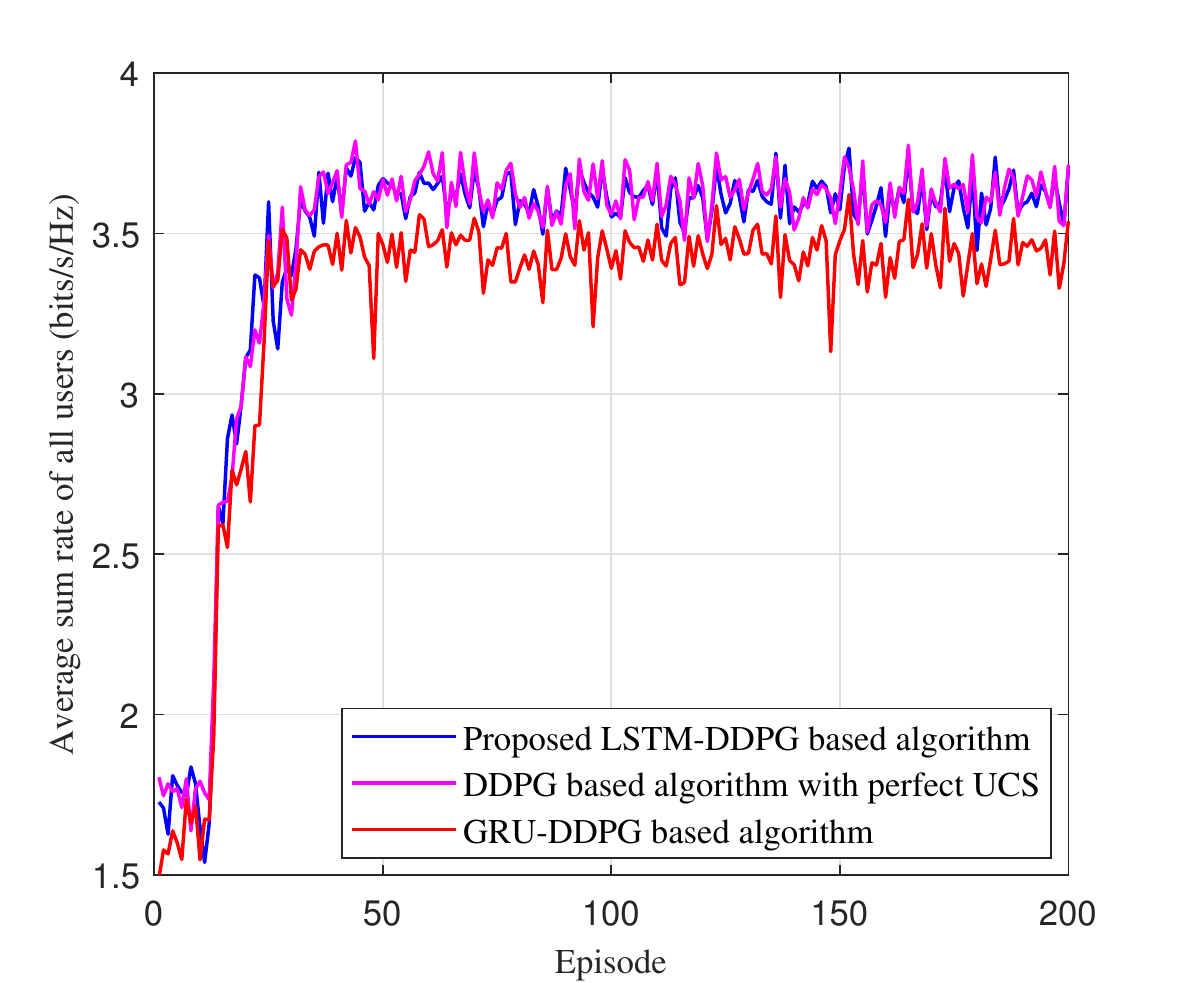}}
	\caption{\textcolor{black}{Performance comparison  with and without LSTM, where  $R_0=0.6$ \text{bits/s/Hz}, $K=4$, $L=25$, $p_k=0.01W$, and $M=100$.} }
	\label{withoutLSTM}
\end{figure}
First, we verify the performance of the proposed LSTM based prediction algorithm. The performance of the LSTM algorithm for a system containing four users is verified in Fig. \ref{LSTM}. The real communication probabilities data of the four users are generated based on a random walk model, and the initial probability for each user is 0.6. Specifically,   a series of 1000 time slots of real data were generated for each user,  and the communication probability of each user in each time slot is a random value in $[0,1]$. These 1000 data will be grouped in 5 time slots to produce 955 data sets, where the communication probability data for every 5th time slot is the input to the LSTM and the label is the communication probability for the 6th time slot. For the 955 datasets generated, 70\% were used for training and the left 30\% for testing. In Fig. \ref{LSTM},  the true data is the communication probability produced by the random model, and the predict data is the estimated output of the LSTM based algorithm. 

As can be seen from Fig. \ref{LSTM}, the LSTM-based prediction algorithm achieves a very high prediction accuracy. \textcolor{black}{The predicted user communication probability and the real communication probability highly overlap,  which proves the convergence of LSTM based prediction algorithm.} The accuracy of the  prediction can provide important guarantees for the subsequent design of the RIS   amplification matrix and phase shift matrix. All subsequent simulation results are obtained based on the LSTM algorithm.

Fig. \ref{withoutLSTM} further validates the performance of the LSTM based prediction algorithm. Specifically, we compare the performance of the proposed LSTM-DDPG algorithm with the DDPG  based  algorithm on perfect UCS information. \textcolor{black}{In addition, to further verify the performance of the proposed LSTM, the gated recurrent unit (GRU) algorithm is introduced in the simulation for comparison, and the  performance of GRU-DDPG based algorithm is verified.} The comparison of the average reward and the average communication success ratio of the system is given in Fig. \ref{withoutLSTM}.  It can be seen that the performance obtained by the proposed LSTM-DDPG algorithm and  the DDPG  based  algorithm with perfect UCS  is almost similar. In the DDPG algorithm, the perfect UCS information does not have a significant advantage over the LSTM based prediction of the UCS information, which fully demonstrates the effectiveness of the designed LSTM algorithm.  \textcolor{black}{In addition, it can be seen from Fig. 5 that the performance of the GRU-DDPG-based algorithm is about 4\% lower than that of the proposed algorithm, and the stability of this algorithm is not good enough.}

\subsection{ \textcolor{black}{Verification of NOMA Performance Advantages}}
\begin{figure}[t]
	\centering
	\subfigure{\includegraphics[width=0.49\textwidth]{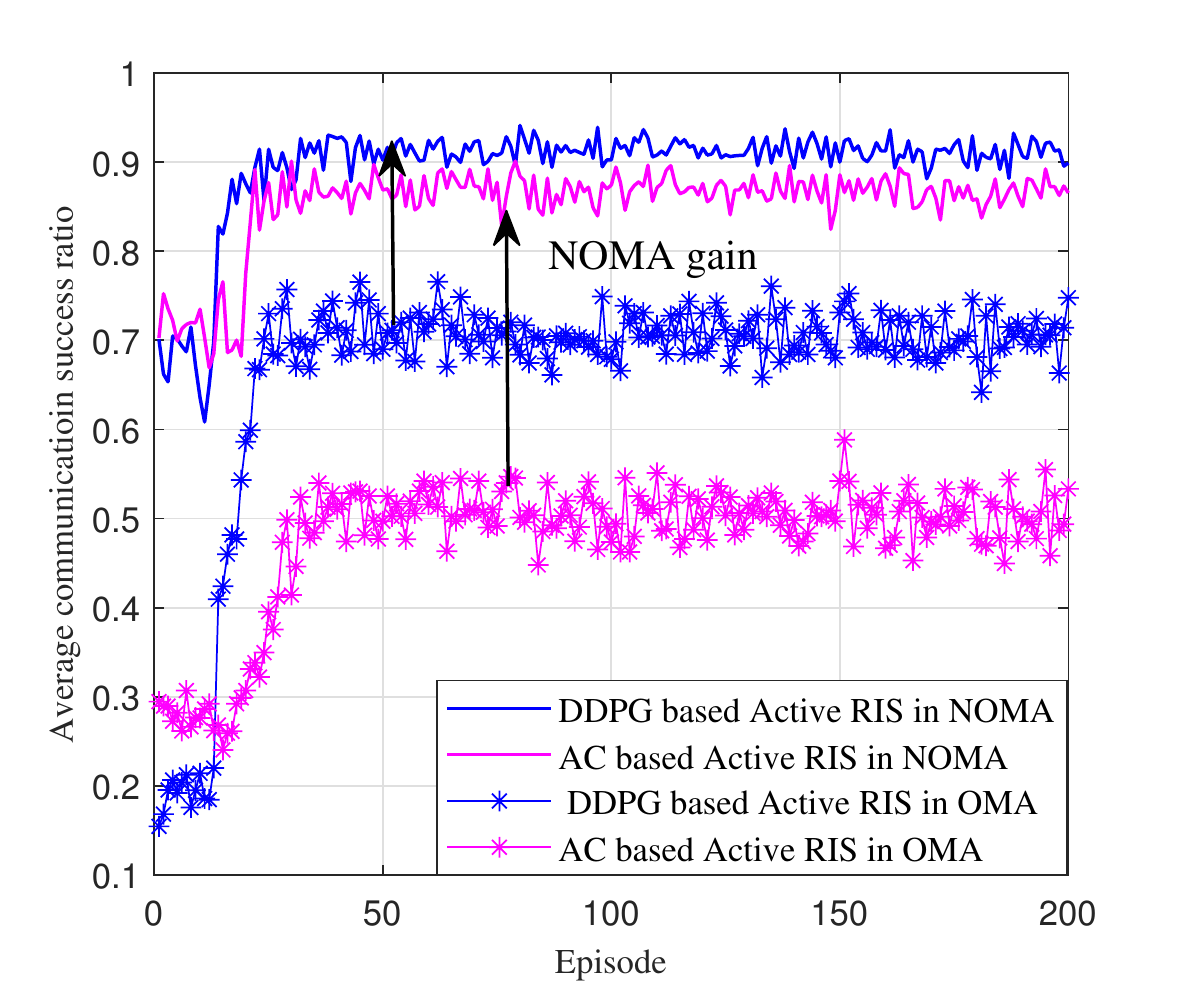}}
	\subfigure{\includegraphics[width=0.49\textwidth]{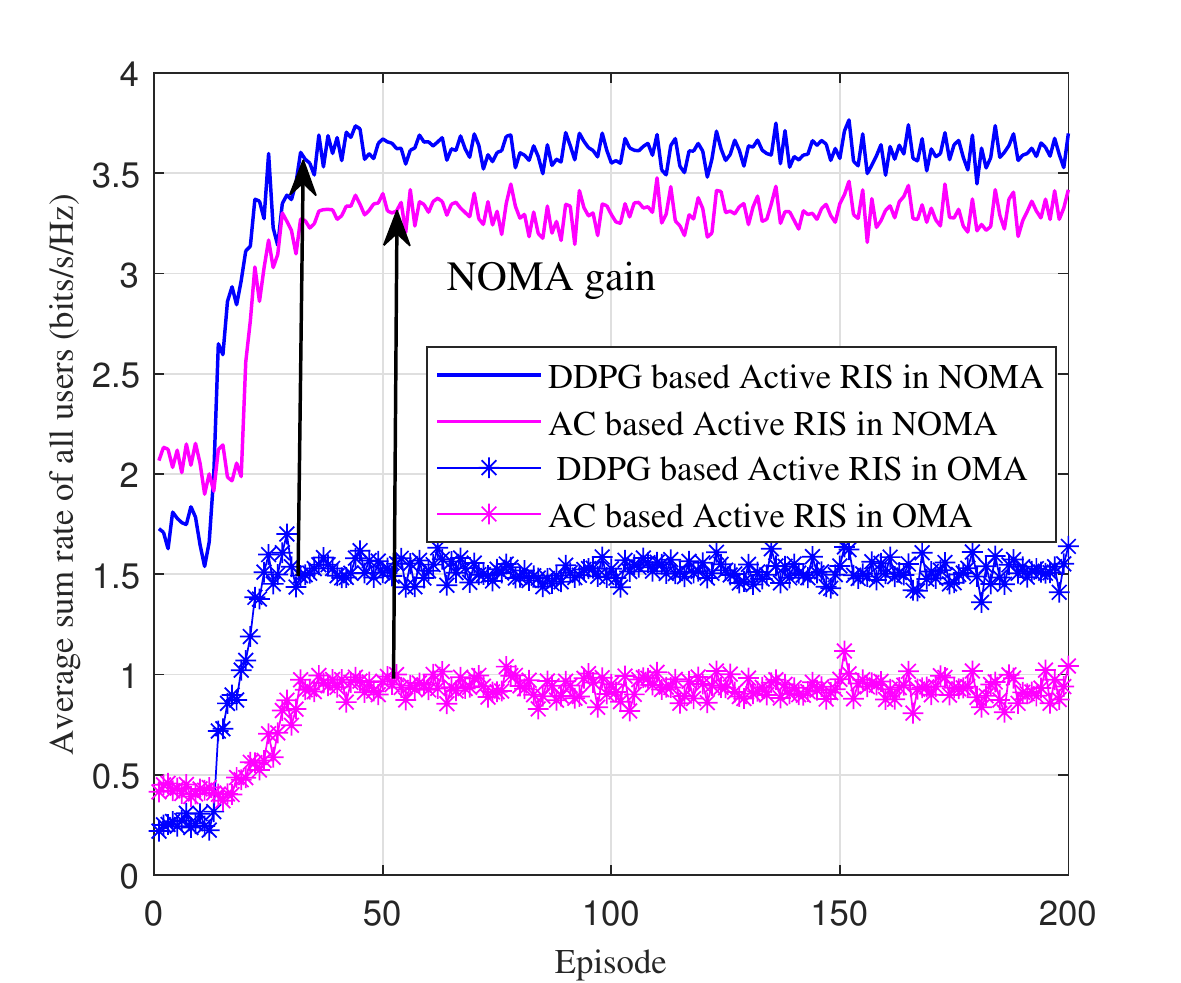}}
	\caption{Performance comparison of in active RIS, where  $R_0=0.6$ \text{bits/s/Hz}, $K=4$, $L=25$,  and $M=100$. }
	\label{active}
\end{figure}
\textcolor{black}{To verify the advantages of using NOMA in  the EH active RIS-aided network, we compared the performance of NOMA and OMA in this network in Fig \ref{active}.}
Two classical DRL algorithms: DDPG and AC are used to perform the design of the amplification power matrix and phase shift matrix of the active RIS. Note that all DRL based algorithms in the simulation use the same  DRL framework, i.e., they  use the same action, state, and reward functions designed in  Section \uppercase\expandafter{\romannumeral4}-A. The training and application results are obtained based on the deep learning framework in TensorFlow 1.14.0.  

First, it can be seen that with training, the performance of both DDPG and AC algorithms can gradually improve and achieve convergence after about 25 episodes in both NOMA and OMA modes, which validates the effectiveness of the proposed DRL framework. 
More importantly,  it is evident that both algorithms are able to obtain better performance in NOMA mode with respect to OMA. \textcolor{black}{The arrows in the diagram visualize the advantages of NOMA technology compared to OMA. Specifically, in the proposed DDPG based algorithm, the average communication success ratio is improved by 29.28\% and the average sum rate  is improved by 137\% in NOMA mode with respect to OMA. The communication success ratio and sum rate gains obtained by NOMA in the AC algorithm can also reach about 72.66\% and 254\%, respectively}. \textcolor{black}{Simulation results show that NOMA can obtain much better performance in active RIS networks compared to OMA, which indicates that NOMA has better adaptability to active RIS networks.}

\begin{figure}[t]
	\centering
	\subfigure{\includegraphics[width=0.49\textwidth]{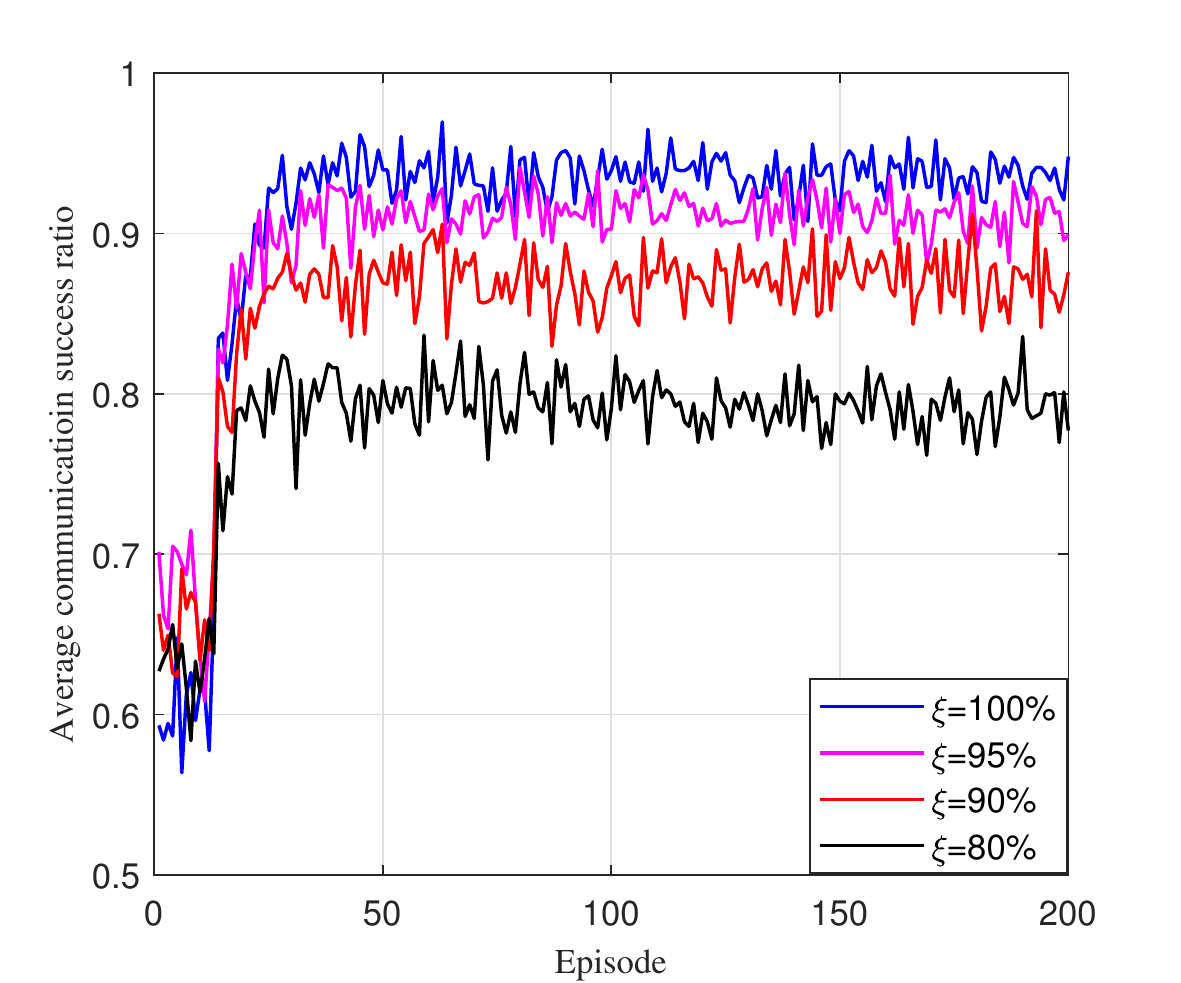}}
	\subfigure{\includegraphics[width=0.49\textwidth]{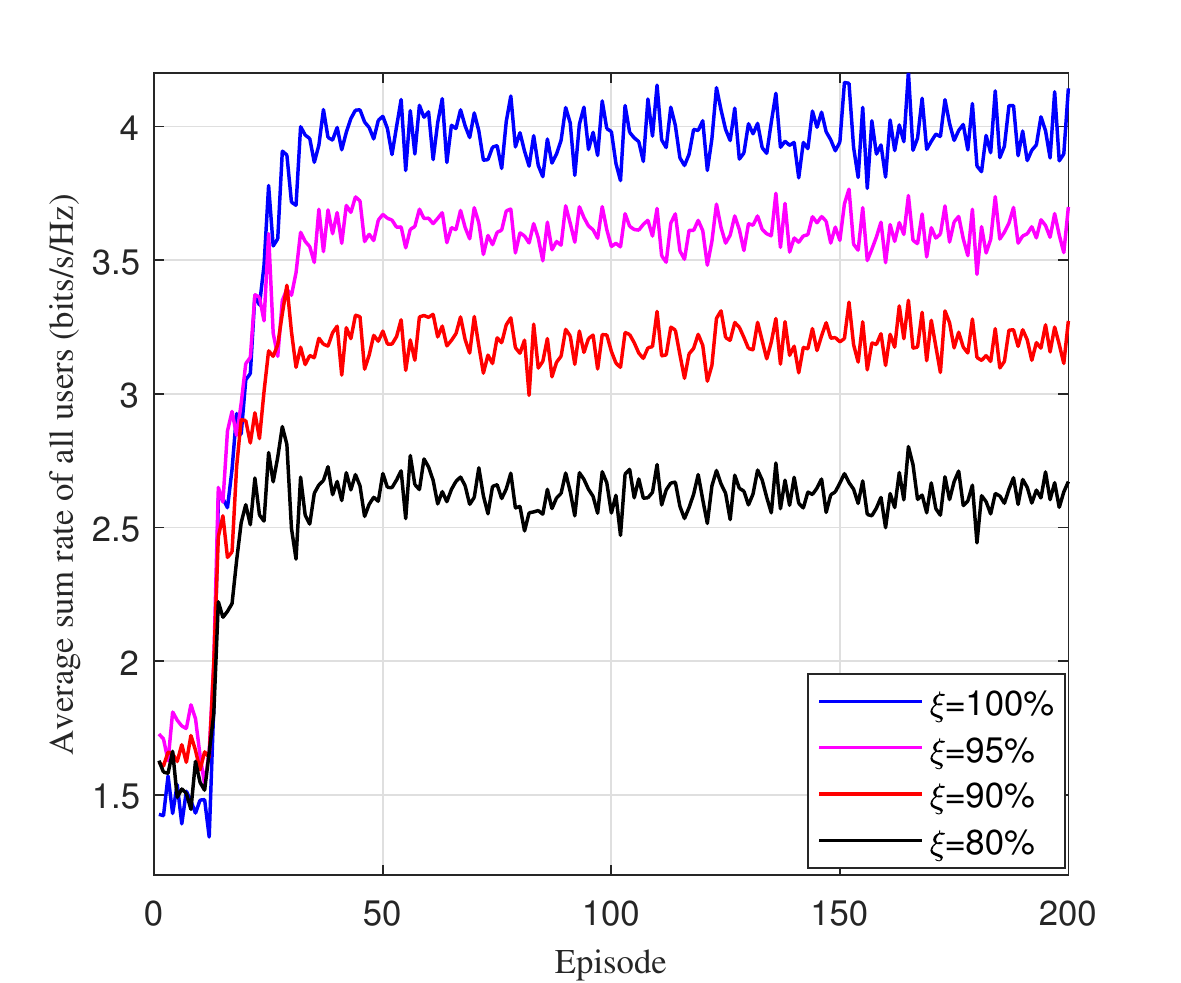}}
	\caption{Impact of the NOMA decoding error $\xi$ on system performance, where  $R_0=0.6$ \text{bits/s/Hz}, $K=4$, $L=25$, and $M=100$. }
	\label{csi}
\end{figure}
\textcolor{black}{Fig. \ref{csi} shows the effect of channel estimation error on the performance of the proposed LSTM-DDPG algorithm. $\xi=100\%$ indicates that the channel estimation error is 0. The smaller the value of $\xi$ characterizes the larger the SIC error of NOMA. It can be seen that the larger the SIC error, the worse the performance obtained by the system, and when $\xi < 90\%$, the average communication success rate of the system is below 0.9.}

\subsection{Performance Comparison of Active RIS  and Passive RIS in the EH-NOMA Networks}
\begin{figure}[t]
	\centering
	\subfigure{\includegraphics[width=0.49\textwidth]{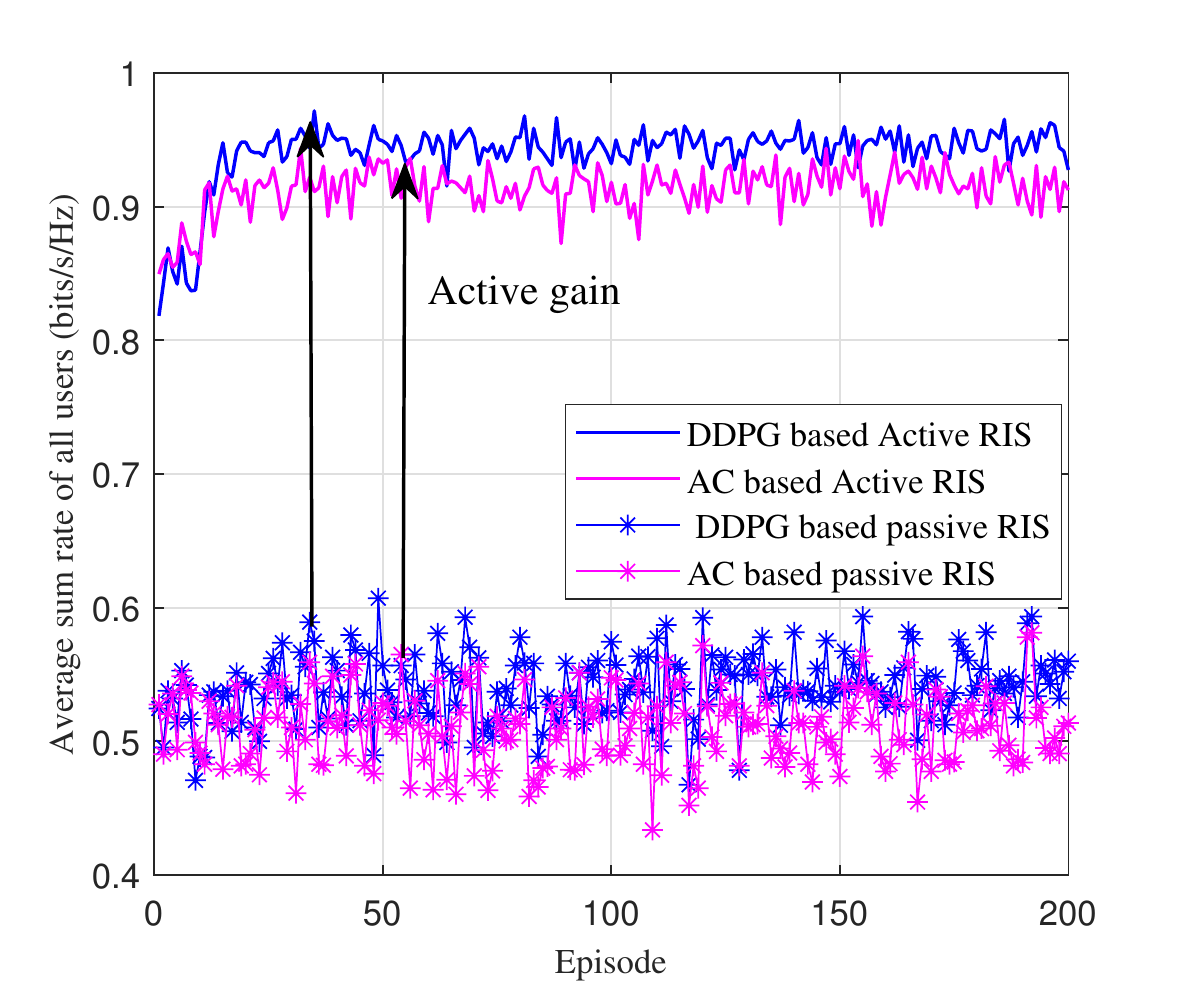}}
	\subfigure{\includegraphics[width=0.49\textwidth]{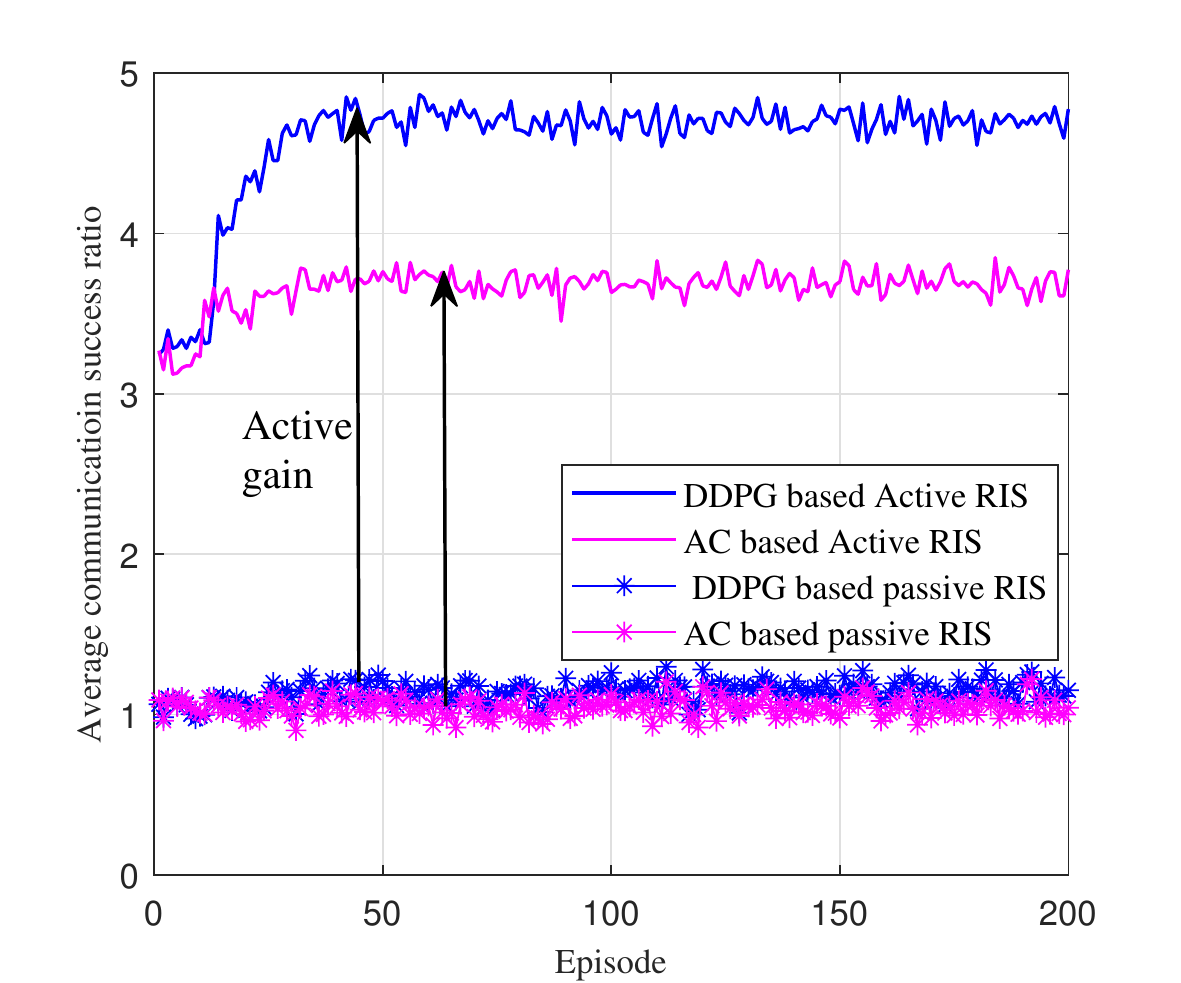}}
	\caption{Performance comparison of active and passive RIS     with NOMA,  where  $R_0=0.6$ \text{bits/s/Hz}, $K=4$, $L=25$, and $M=100$. }
	\label{passive}
\end{figure}

Fig. \ref{passive} shows the performance of the two DRL algorithms, DDPG and AC, under active and passive RIS, respectively. \textcolor{black}{Note that since the system performance is very poor (below the average communication success rate of 0.1) in the passive RIS system when the user transmit power is not high ($p_k=0.01 W$), the transmit power of the user is set to $p_k=0.05W$ in all subsequent simulations for better performance comparison.} All the users in the simulation are accessing the network by NOMA.  As shown in Fig. \ref{passive}, the control of the phase matrix by the two DRL algorithms do not achieve any performance improvement under passive RIS, so both algorithms obtain similar performance. However, under active RIS, both DRL algorithms can achieve a significant performance improvement after a short training period. This is because the amplification matrix and the phase shifting matrix are jointly controlled in the active RIS.  

The advantages of active RIS   over passive RIS    can be clearly seen in Fig. \ref{passive}. \textcolor{black}{Specifically, in the proposed DDPG algorithm, the communication success ratio and the sum rate performance of the active RIS   are improved by 73.72\% and 306.3\%, respectively, compared to the passive RIS. The active RIS   in the AC algorithm also obtains 79.3\% and 251\% performance gains in these two performances, respectively}. Last but not least, both Figs. \ref{active} and  \ref{passive} show that the DDPG algorithm achieves better training results than the AC algorithm.  This is because the DDPG algorithm introduces the structure of DQN to improve the stability and convergence of AC.


\subsection{ Application Performance Testing of the Proposed LSTM-DDPG  Algorithm }
\begin{figure}[t]
	\centering
	\subfigure[]{\includegraphics[width=0.45\textwidth]{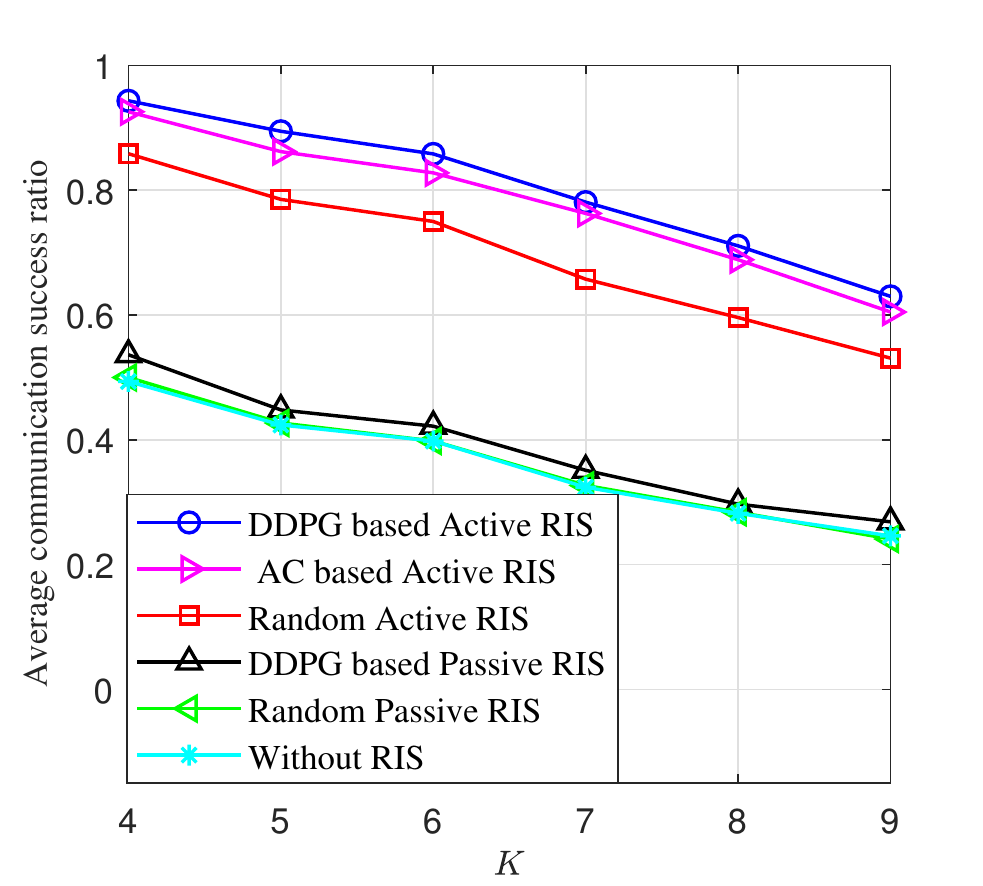}}
	\subfigure[ ]{\includegraphics[width=0.45\textwidth]{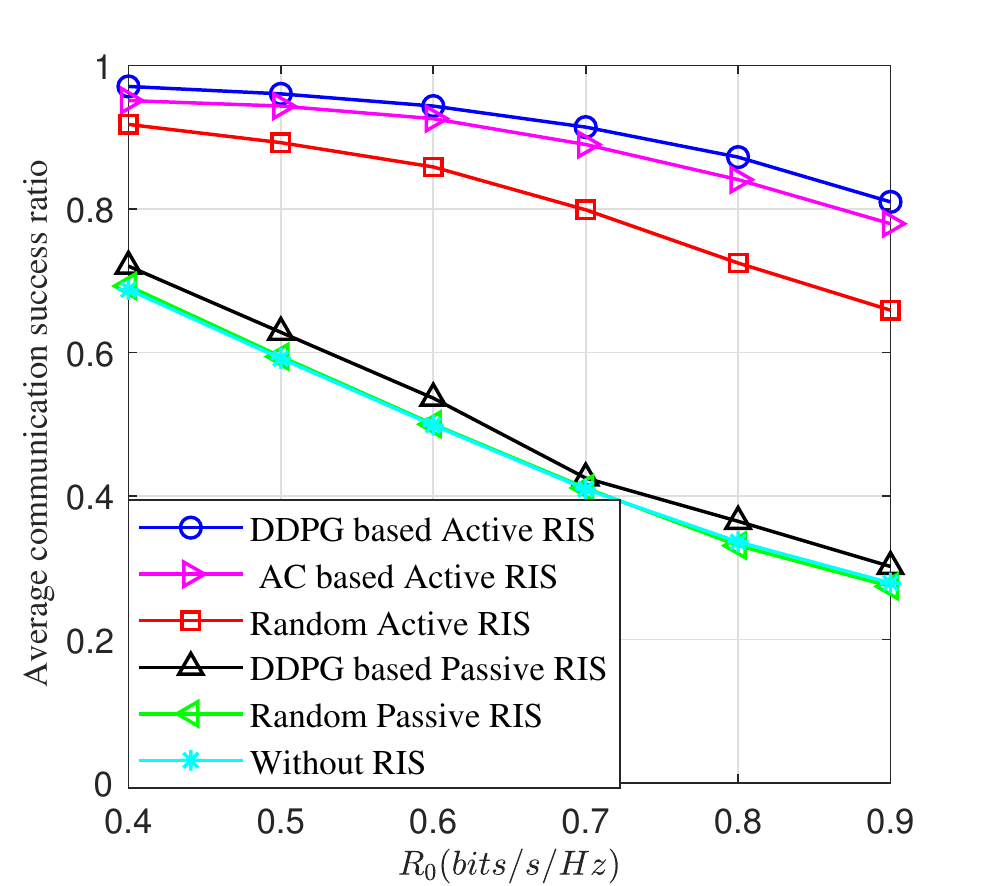}}
	\caption{Application performance comparison of different algorithms, where   $L=25$  and $M=100$.}
	\label{N}
\end{figure}

\begin{figure}[t]
	\centering
	\subfigure[]{\includegraphics[width=0.45\textwidth]{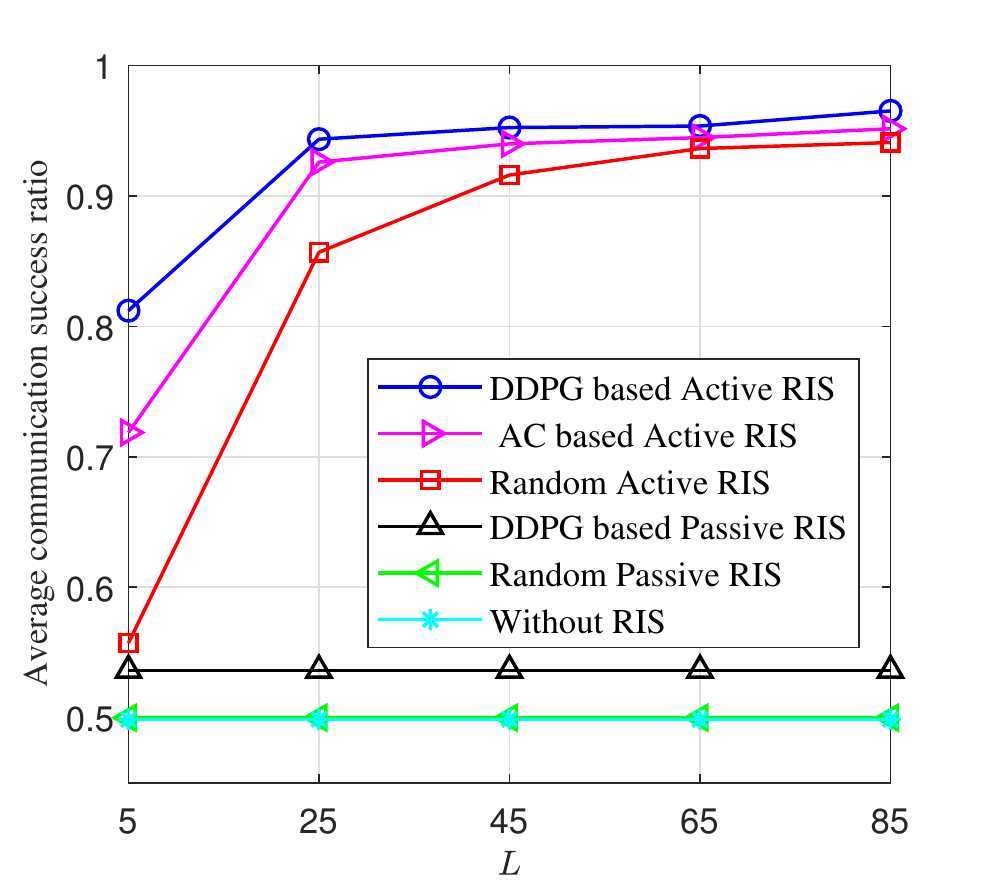}}
	\subfigure[ ]{\includegraphics[width=0.45\textwidth]{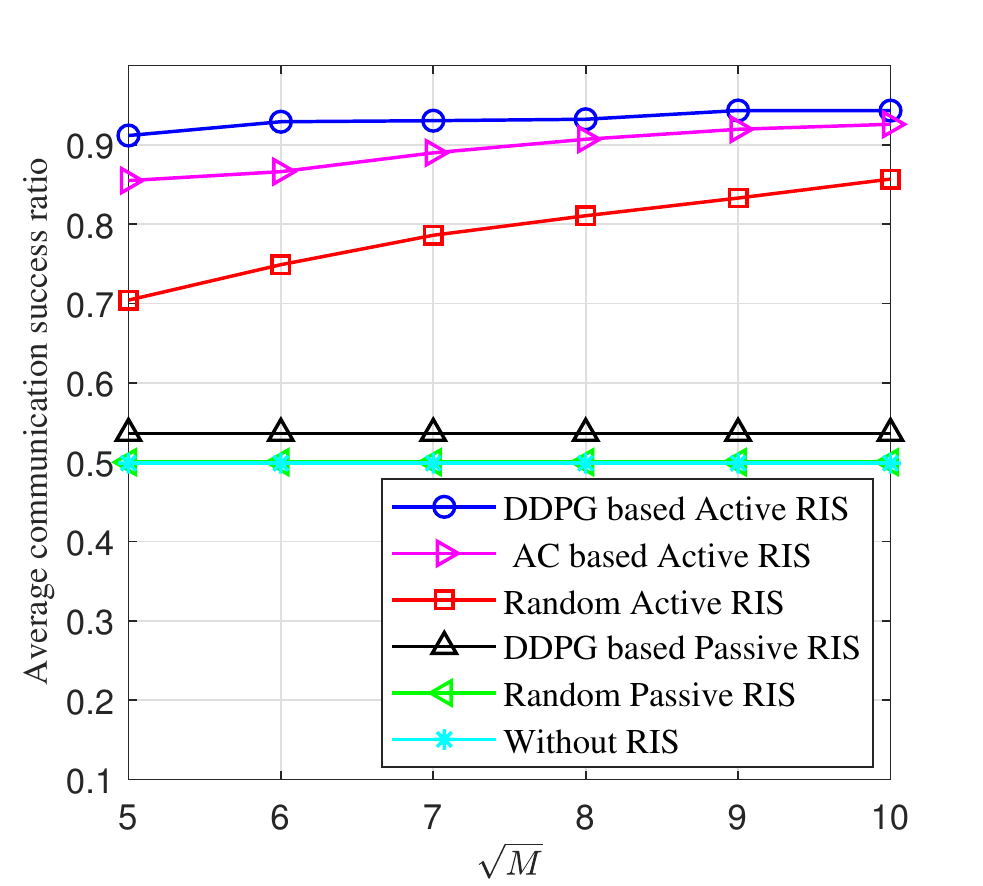}}
	\caption{Application performance comparison of different algorithms, where   $R_0=0.6 bits/s/Hz$ and $K=4$.}
	\label{M}
\end{figure}
Finally, the application performance of the proposed LSTM-DDPG algorithm is tested. Considering that the optimization objective of our work is to maximize the communication success ratio of the whole system, we test the ratio of the trained algorithm. The data for training and testing are generated simultaneously and distributed in a $7:3$ ratio.  To better demonstrate the superiority of the proposed algorithm, several benchmark algorithms are introduced for comparison  via different system parameters in Figs. \ref{N} - \ref{M}.  It can be seen that the performance of the passive RIS network and the without RIS  network is similarly poor under various network parameters, and the networks without RIS  have the worst performance. This is because there is no power amplification in these networks, thus failing to meet the required rate threshold and causing the communication to fail.

In Fig. \ref{N} (a), we examined the impact of the number of users $K$ on the system performance of different algorithms with $R_0=0.6$ \text{bits/s/Hz}. It can be seen that the performance of all the algorithms gets worse as $K$ increases. This is due to the fact that more co-channel interference will be caused by more users, which makes lower rates, and thus more users are unable to communicate successfully. It can be seen that the two DRL algorithms can obtain better performance compared to other algorithms, with the DDPG based algorithm obtaining the best communication success ratio. Although both algorithm apply active RIS, the random active RIS   algorithm performs worse than the DRL based algorithm. This is because the amplification matrix and phase shift matrix in the random active RIS   algorithm are not determined based on the current system state, but are determined randomly. In addition, it can be found that the proposed DDPG based algorithm can obtain better performance than other algorithms. \textcolor{black}{Even when $K=9$, the DDPG based algorithm still achieves the performance of about $0.63$, outperforming the AC, random active RIS,  DDPG based passive RIS, and random passive RIS algorithms about by 5\%, 18.9\%, 135.1\%, and 156.1\%, respectively.}



Fig. \ref{N}(b) illustrates the impact of the rate threshold on all comparison algorithms with $K=4$. As shown in Fig. \ref{N}(b), the performance of all algorithms deteriorates as $R_0$ increases. This is because a higher rate threshold means it is more challenging to achieve successful communication, which leads to more task failures.  \textcolor{black}{Also, it can be seen that even at $R_0 = 0.9$, the DDPG algorithm still obtains a communication success ratio of 0.81, which is 3.8\%, 22.7\%, and 170\%   better than the AC, the random active RIS, and the DDPG based passive RIS, respectively.}

Fig. \ref{M} examines the effect of the maximum power amplification $L$ and the number of RIS   elements $M$ on the performance of different algorithms, where  $M=100$ in Fig. \ref{M}(a), $L=25$ in Fig. \ref{M}(b). It can be seen that the performance of the two passive RIS   algorithms as well as the without RIS   network is not affected by the parameters $L$ and $M$. This is because in all three algorithms, the RIS   does not amplify the received signal, and the increase in the number of RIS   elements does not improve the user's communication success when the rate threshold $R_0$ is not very small ($R_0$ = 0.6 bits/s/Hz).  In addition, it can be observed that the communication success ratio of the three active RIS   algorithms improves significantly with increasing $L$ and $M$, especially for the two DRL based algorithms, which further supports the validity of the Markov model of the designed DRL.\textcolor{black}{ Among all the algorithms, the proposed DDPG algorithm is able to achieve the best performance. Specifically,  the proposed algorithm  achieves a communication success ratio of 0.943 with $L = 100$ and $M = 100$, outperforming the AC, the random active RIS, DDPG based passive RIS, and random passive  algorithm by  1.86\%, 10.07\%, 75.6\%, and 88.6\%, respectively.}


\section{Conclusions}

In this paper, we studied the RIS   control for active RIS-aided EH-NOMA networks, where the RIS is powered by EH, and the UCS are dynamic. With the objective of maximizing the communication success ratio, an LSMT based algorithm was designed to predict the UCS. Based on the prediction results, a DDPG based algorithm was proposed for the joint control of the amplification matrix and phase shift matrix of the active RIS. The complexity of the LSTM based prediction algorithm and the DDPG based RIS control algorithm are analyzed.  Sufficient simulation results demonstrated the effectiveness and superiority of the proposed algorithm in terms of the communication success ratio. \textcolor{black}{Simulation results demonstrate the achievable performance improvement for our proposed scheme with respect to schemes of OMA,  NOMA with passive RIS, and  NOMA without RIS.} In the future, we will consider the power allocation at the BS and further explore RIS  control algorithms for more complex communication scenarios, such as systems containing multiple channels and (or) multiple antennas configured at the BS or users.
\appendices

\end{document}